\documentclass[10pt,journal,compsoc]{IEEEtran}

\usepackage{amsmath,amssymb,amsfonts}
\usepackage{algorithmic}
\usepackage{graphicx}
\usepackage{textcomp}
\usepackage{xcolor}
\usepackage[caption=false]{subfig}
\usepackage{float}
\usepackage{bm}
\usepackage{soul}
\usepackage{comment}
\usepackage{subfig}
\usepackage{bm}
\usepackage{ragged2e}
\usepackage{amsmath}
\usepackage{xspace}
\usepackage{enumitem}
\usepackage{url}
\usepackage{breakurl}
\usepackage{dblfloatfix}
\usepackage{balance}

\DeclareMathOperator*{\argmin}{arg\,min}

\ifCLASSOPTIONcompsoc
  \usepackage[nocompress]{cite}
\else
  \usepackage{cite}
\fi

\ifCLASSINFOpdf

\else

\fi

\ifodd 1
\else

\fi

\newcommand{\name}{RayLoc\xspace}
\begin{document}

\title{\name: Wireless Indoor Localization via Fully Differentiable Ray-tracing}

\author{Xueqiang Han,~\IEEEmembership{Student Member,~IEEE},
        Tianyue Zheng,~\IEEEmembership{Member,~IEEE},
        Tony Xiao Han,~\IEEEmembership{Senior Member,~IEEE},
        and Jun Luo,~\IEEEmembership{Fellow,~IEEE} %
\IEEEcompsocitemizethanks{

\IEEEcompsocthanksitem X. Han and T. Zheng are with the Department of Computer Science and Engineering, Southern University of Science and Technology, China. \protect\\
E-mail: \{hanxq, zhengty\}@sustech.edu.cn
\IEEEcompsocthanksitem T. X. Han is with Huawei Technologies Co., Ltd, China. \protect\\
E-mail: tony.hanxiao@huawei.com
\IEEEcompsocthanksitem J. Luo is with the College of Computing and Data Science, Nanyang Technological University, Singapore.\protect\\
E-mail: junluo@ntu.edu.sg}
}

\IEEEtitleabstractindextext{%
\begin{abstract}
\justifying 
Wireless indoor localization has been a pivotal area of research over the last two decades, becoming a cornerstone for numerous sensing applications.  However, conventional wireless localization methods rely on channel state information to perform blind modelling and estimation of a limited set of localization parameters. This oversimplification neglects many  sensing scene details, resulting in suboptimal localization accuracy. To address this limitation, this paper presents a novel approach to wireless indoor localization by reformulating it as an inverse problem of wireless ray-tracing, inferring scene parameters that generates the measured CSI. At the core of our solution is a fully differentiable ray-tracing simulator that enables backpropagation to comprehensive parameters of the sensing scene, allowing for precise localization. To establish a robust localization context, \name constructs a high-fidelity sensing scene by refining coarse-grained background model. Furthermore, \name overcomes the challenges of sparse gradient and local minima by convolving the signal generation process with a Gaussian kernel. Extensive experiments showcase that \name outperforms traditional localization baselines and is able to generalize to different sensing environments.

\end{abstract}

}

\maketitle

\IEEEdisplaynontitleabstractindextext

\IEEEpeerreviewmaketitle

\IEEEraisesectionheading{\section{Introduction}\label{sec:introduction}}
Location-based services have become ubiquitous in modern life, driving extensive research in localization techniques across both academia~\cite{lbs}
and industry~\cite{google-map, cisco-space, iBeacon}. While Global Navigation Satellite Systems (GNSS) dominate outdoor positioning, wireless indoor localization has emerged as its essential counterpart, addressing the crucial need for positioning in indoor environments where GNSS signals cannot penetrate~\cite{survey}. Drawing a parallel to how GNSS utilizes satellites as celestial reference points, indoor localization systems leverage indoor wireless infrastructure to determine the precise location of devices and targets. These systems have found widespread adoption across numerous applications, ranging from building management~\cite{survey} and smart homes~\cite{smart-home} to robot navigation~\cite{dloc} and security surveillance~\cite{surveillance}. However, as these applications grow increasingly sophisticated, they demand ever-higher levels of localization accuracy. %

Early wireless indoor localization systems rely on coarse-grained received signal strength (RSS) from multiple access points (APs) to determine target positions~\cite{rss-radar, pain-loc, rass, rss-tmc-loc}, yielding relatively imprecise localization results. Recent advances have introduced more sophisticated approaches utilizing fine-grained channel state information (CSI)\cite{FILA, spotfi, ArrayTrack} and beamforming feedback information (BFI, a compressed version of CSI)\cite{hu2023muse, BFMSense} for improved localization accuracy. However, these methods can only model and estimate a limited set of localization parameters, such as time of flight (ToF), angle of arrival (AoA), and angle of departure (AoD). The oversimplification inherent in these parameter-constrained approaches can neglect details and misrepresent the sensing scene, particularly when dealing with targets that cannot be adequately represented as point reflectors. Moreover, hardware limitations in bandwidth and antenna array size often result in poor range and angular resolution, further compromising localization accuracy.

Several methods enhance localization accuracy by explicitly modeling and estimating sensing scenes: iLocScan~\cite{loc-acce} combines prior knowledge of wall configurations and super-resolution algorithms to simultaneously perform indoor source localization and space scanning. Similarly, mD-Track~\cite{xie2019md} employs an iterative path separation algorithm to distinguish signals reflected from targets of interest from those reflected by background structures, enabling estimation of Angle of Arrival (AoA) and Time of Flight (ToF) for multipath components. $M^3$~\cite{m3} takes a different approach, leveraging the knowledge of nearby reflectors to achieve sub-meter localization using just a single Access Point (AP). However, these methods share a fundamental limitation: their modeling and estimation of the sensing scene remain coarse-grained: they attempt to represent complex real-world sensing scenes with simplistic reflectors at previously measured or coarsely estimated locations, an approach that is inherently under-parameterized. 

To address this under-parameterization challenge, researchers have turned to deep learning (DL) networks, which offer vastly more parameters to implicitly represent the sensing scene~\cite{Thz-loc, will, NNE}. These approaches achieve sub-meter-level accuracy by employing data-driven methods to train location classifiers or regressors that map wireless signal fingerprints to object locations based on wireless propagation characteristics. Despite their claimed impressive accuracy, these data-driven approaches suffer from two critical limitations: they are highly susceptible to even minor variations of scene parameters~\cite{DAFI}, and their black-box nature, specifically, the opacity of how they encode sensing scene parameters within their neural network architectures, raises serious concerns about their ability to generalize to unseen environments.  %

\begin{figure}[t] %
    \centering %
    \includegraphics[width=0.48\textwidth]{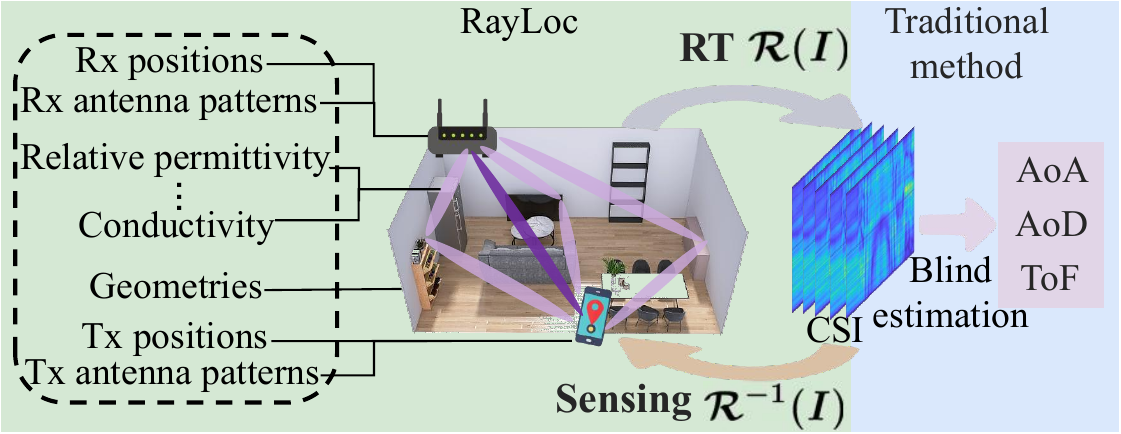} %
    \caption{Unlike traditional wireless localization performing blind estimation of a few parameters based on CSI, \name infers the full-set scene parameters by the inverse process of differentiable wireless ray-tracing.} %
    \label{fig:inverse-loc} %
\end{figure}

To fully address the issues of under-parameterization in scene parameter modeling and estimation, researchers have recently made advances in developing differentiable digital replicas of wireless sensing environments~\cite{winert,nerf2,Sionna}. These replicas enable precise optimization of various physical properties through end-to-end differentiable simulations. A notable example is Sionna~\cite{Sionna}, which implements differentiable RT simulations by modeling physical interactions within the sensing scene using differentiable building blocks. However, Sionna does not support full differentiability with respect to target locations, forbidding inferring locations using the inverse process. Alternative DL-based RT approaches such as WiNeRT~\cite{winert} and Nerf$^2$~\cite{nerf2} approximate the sensing environment as a DL network. While differentiable, these methods suffer from fundamental limitations: their lack of explainability prevents tracing specific network parameters back to physical target locations. Moreover, as mentioned previously, their black-box nature also raises generalization concerns. %

By leveraging a fully parameterized model of the sensing environment, where target positions are key parameters, we can reframe wireless indoor localization as an inverse problem of RT, offering a novel path to precise localization. As such, we propose a differentiable wireless RT simulation system that models full-scale scene parameters and accurately simulates wireless signal propagation to generate corresponding CSIs. The localization process is achieved through gradient-based optimization of target-specific parameters, which minimizes the difference between simulated and observed CSIs. What sets differentiable RT methods apart is their unique ability to disentangle the interactions of multiple scene parameters. This capability represents a fundamental paradigm shift: environment parameters treated as noise in conventional localization methods are instead harnessed as valuable, information-rich signals. %

However, achieving wireless indoor localization by solving the inverse problem of wireless simulation presents three significant challenges. First, to the best of our knowledge, no fully differentiable RT-based wireless simulator capable of enabling gradient calculation with respect to scene geometries has been proposed.  Second, constructing a high-fidelity virtual model of the physical world remains highly non-trivial. Even intricate discrepancies between the virtual and physical models must be resolved to enable high-precision CSI simulation. Lastly, even if a fully differentiable RT-based simulator is available to provide gradients for object positions, it would still face inherent optimization challenges. Specifically, the sparse gradients and local minima in the loss landscape resulting from the limited spatial footprint of the optimized target hinder effective optimization, making it challenging to converge on the correct target location.

To address these challenges, we present \name, a wireless indoor localization system that leverages the inverse process of differentiable RT. First, we design a fully differentiable wireless ray tracer that backpropagates to object positions. By creating a virtual model of the physical sensing environment, the differentiable RT engine enables us to estimate the target position through gradient-based optimization, minimizing the discrepancy between simulated and real-world reference CSIs. To construct an accurate virtual model of the physical environment, containing both geometry and material electromagnetic (EM) properties, that feeds the RT, \name first constructs a high-fidelity background model by refining a coarse-grained model provided by the floor plan. Finally, to address the challenge of sparse gradients and local minima in the loss landscape, \name incorporates a Gaussian kernel to smooth the loss landscape associated with object positions during CSI generation, enhancing convergence and improving localization accuracy. In summary, the key contributions of this paper are as follows:

\begin{itemize}
    \item To the best of our knowledge, \name is the first wireless indoor localization system that unifies both \emph{device-free} and \emph{device-based} localization leveraging the inverse process of differentiable RT.
    \item We design a fully differentiable RT-based wireless simulator that enables gradient descent-based localization, while existing such as like Sionna~\cite{sionna-rt} lacks full differentiability with respect to target locations. 
    \item To achieve high-precision CSI simulation, we introduce a novel data-driven  method that constucts high-fidelity model of the sensing scene. 
    \item We employ a convolution with Gaussian kernel to smooth the loss landscape, alleviating sparse gradient and local minima, thus improving convergence.
    \item We implement the \name prototype and conduct extensive experiments, demonstrating its ability to achieve highly accurate localization with remarkable generalization across diverse environments.
\end{itemize}

The rest of this paper is structured as follows. \S~\ref{sec:bm} introduces the background and motivation of our work. \S~\ref{sec:design} presents the system design of \name. The implementation detail and extensive experiments are discussed in \S~\ref{sec:impl} and \S~\ref{sec:evaluation}, respectively. \S~\ref{sec:discussion} presents the discussion of potential future work. Finally, we conclude our paper in \S~\ref{sec:conclusion}.
\section{Background and Motivation}\label{sec:bm}
In this section, we introduce the preliminaries of RT and investigate the impact of variation in scene parameters on CSI. We subsequently show that the sparse gradient region and local minima exist in the loss landscape under device-free and device-based configurations.

\subsection{Preliminaries of Wireless RT}\label{moti-grad}
RT provides an efficient geometric approach to model EM wave propagation by simulating discrete rays between a transmitter (Tx) and receiver (Rx), accounting for key propagation phenomena such as reflection, scattering, and diffraction through interaction with the environment. This method has been widely recognized as a key enabler for various applications~\cite{beamforming, uav-loc}. Notably, RT can closely approximate solutions to Maxwell equations, which govern the underlying wave physics, while remaining computationally tractable even in complex scenarios where solving the full wave equations becomes impractical.
Given the scene parameters $\mathbf{I}$, including the geometries, physical properties of object material, along with the transceiver parameters, RT can render multiple propagation paths of wireless signals and produce corresponding CSIs $\mathbf{H}$. Formally, the RT simulation process can be interpreted as a function $\mathbf{H} = \mathcal{R}(\mathbf{I})$. In this paper, we focus on CSI as the state-of-the-art wireless indoor localization approaches all rely on CSI for accurate localization~\cite{FILA, spotfi, m3, ArrayTrack, DAFI}.

If we could reverse the RT simulation process $\mathcal{R}$, we would be able to estimate the scene parameters $\mathbf{I}$ directly from the observed channel state information (CSI). While a direct inverse function $\mathcal{R}^{-1}(\mathbf{H})$ is intractable due to the complexity of wireless signal interactions, we can formulate this inverse problem as an optimization task. This approach requires two key components: first, a loss function $L=g(\mathbf{H})$ that quantifies the discrepancy between real-world CSI observations and their simulated counterparts, where the gradient $\frac{\partial L}{\partial \mathbf{I}}$ guides the optimization of scene parameters; and second, a differentiable forward RT simulation process, which will be detailed in Section~\ref{ssec:fully_diff}. This optimization-based inverse process elegantly unifies what has traditionally been treated as two distinct sensing paradigms: \emph{device-free}~\cite{widar2} and \emph{device-based} localization~\cite{spotfi}, as both object and transceiver parameters are inherently part the scene parameters $\mathbf{I}$, and we will demonstrate the unified solution and its result in Section~\ref{ssec:fully_diff} and~\ref{sec:evaluation}, respectively.

\subsection{Inaccuracies of Scene Parameters}%
The accuracy of RT simulations is heavily dependent on scene parameters that are challenging to measure precisely in real-world environments. These parameters include both the exact localization of objects and their EM material properties, such as conductivity and relative permittivity. While floor plans can provide basic geometries and EM material properties, objects in physical environments often deviate from the coarse-grained models by several centimeters. Additionally, EM properties of the material measured in controlled settings, whether through empirical testing or vector network analyzer measurements, frequently differ from their real-world behavior. These inaccuracies in both geometry and EM properties of the material can significantly impact the accuracy of wireless signal simulations.

\begin{figure}[htb]
	\captionsetup[subfigure]{justification=centering}
		\centering
		\subfloat[Location inaccuracies.]{
        \hspace{-2ex}
		  \begin{minipage}[b]{0.49\linewidth}
		        \centering
			    \includegraphics[width = \textwidth]{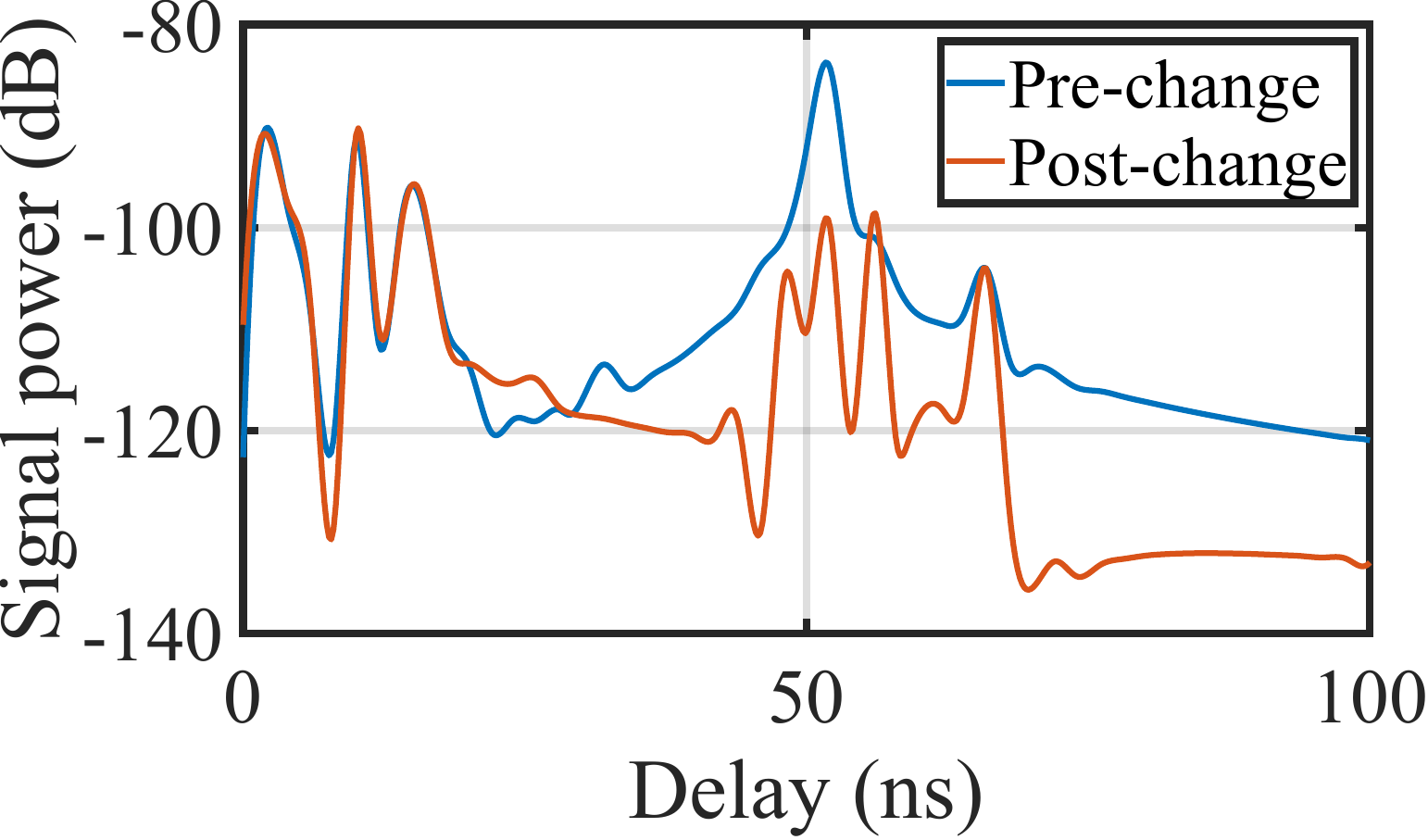}                   
			\end{minipage}
			\label{fig:geo-pow}
		}
		\subfloat[EM property inaccuracies.]{
		    \begin{minipage}[b]{0.49\linewidth}
		        \centering
			    \includegraphics[width = \textwidth]{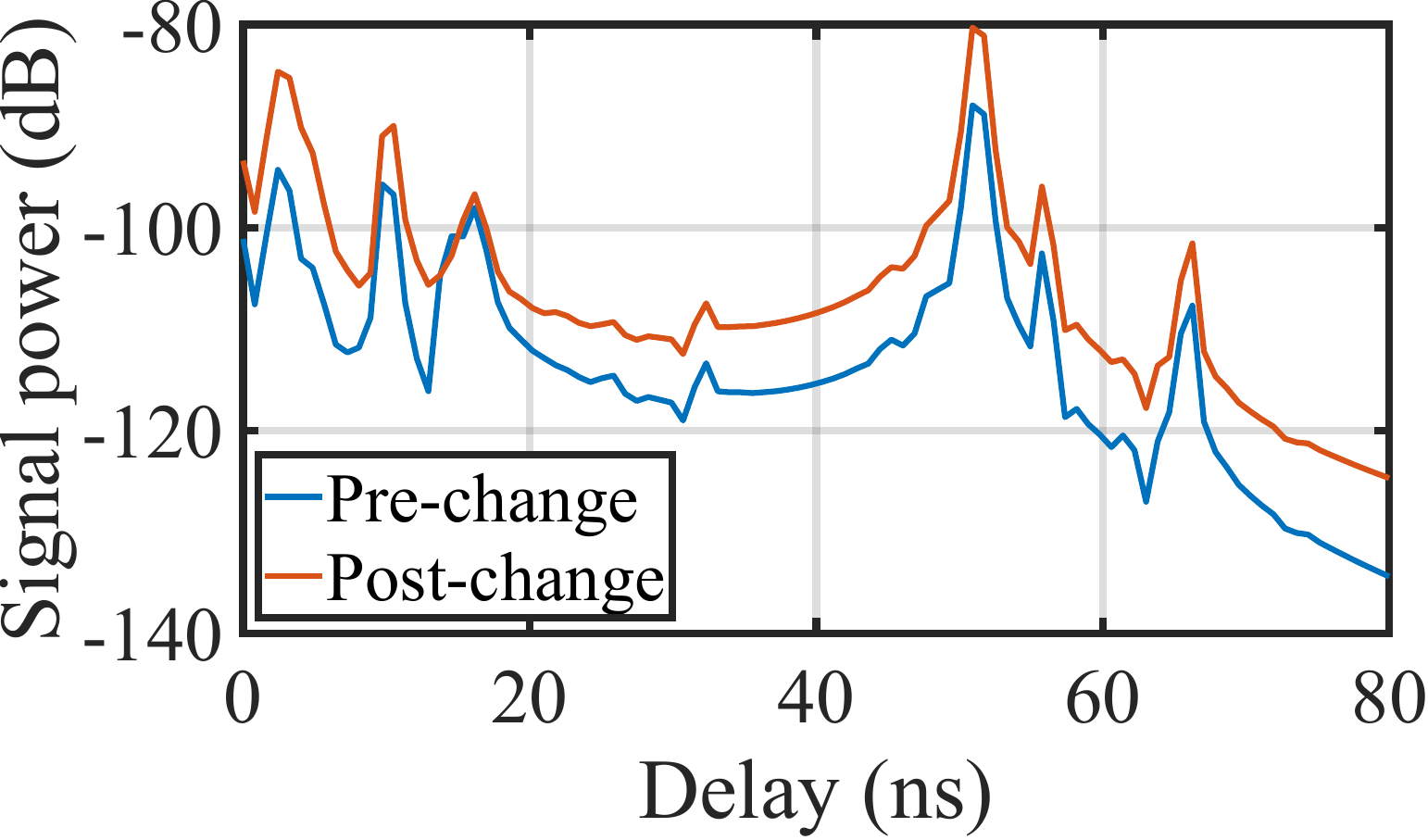}
                    
			\end{minipage}
			\label{}
		}
		\caption{Discrepancies between pre-change and post-change power delay profiles arise from inaccuracies in locations and EM properties.}
		\label{fig:moti-scene-par}	
\end{figure}

To demonstrate how inaccuracies in scene parameters affect wireless signal propagation, we model a typical indoor office environment and conduct experiments in the scene. Specifically, we use a Tx with vertically polarized antenna and an Rx with one antenna, both positioned 1~\!m above ground level, to acquire the CSI. We then shift a sofa 5~\!cm along the $y$-axis from its true position. For material EM property assessment, we maintain object positions while varying the floor's relative permittivity from 5.24 (ITU-R P.2040-2~\cite{itu} standard for concrete) to 10.0 (typical for wet concrete~\cite{concret-wet}).
After performing RT simulations and measuring the CSI, we convert the frequency-domain CSI to time-domain power-delay profile using inverse fast Fourier transformation (IFFT)\cite{book-wireless}. Our findings in Fig.~\ref{fig:moti-scene-par} demonstrate that even minimal spatial displacement (only centimeters) significantly alters wireless signal propagation patterns. Similarly, changes in the floor's EM properties substantially affect energy distribution across all paths. These results demonstrate that accurate RT simulation fundamentally depends on the construction of detailed, high-fidelity sensing scenes, forbiding the direct use of coarse-grained scene models (e.g., those provided by floor plans). %

\subsection{Optimization-Unfriendly Loss Landscape}\label{sec:plateaus}
\begin{figure}[h]
	\captionsetup[subfigure]{justification=centering}
		\centering
		\subfloat[The scene layout and process of target position change.]{
		  \begin{minipage}[b]{0.49\linewidth}
		        \centering
			    \includegraphics[width = \textwidth, height = 0.62\textwidth]{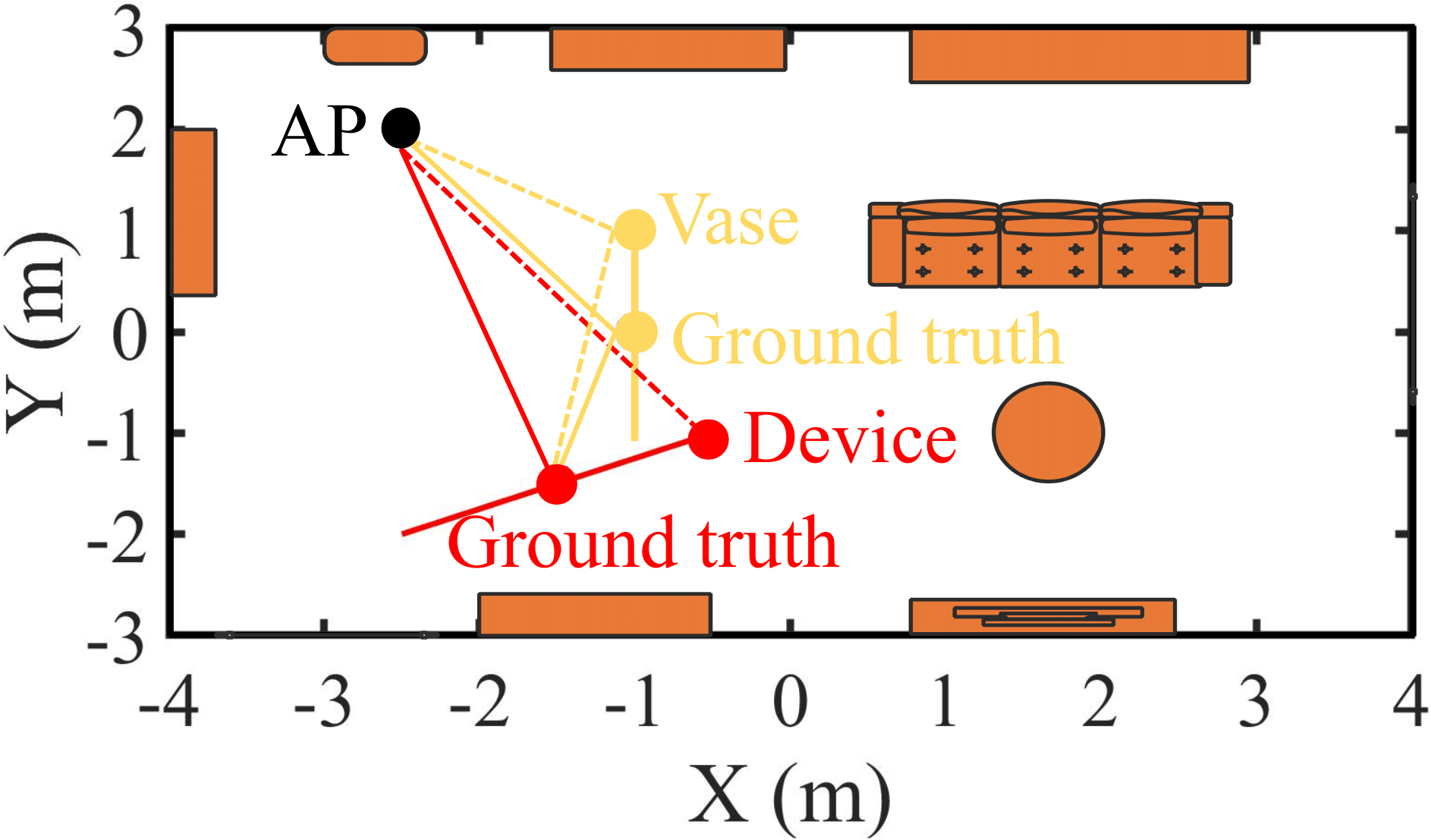}                   
			\end{minipage}
			\label{fig:moti-scene}
		}
		\subfloat[The losses in device-free and device-based settings.]{
		    \begin{minipage}[b]{0.49\linewidth}
		        \centering
			    \includegraphics[width = \textwidth]{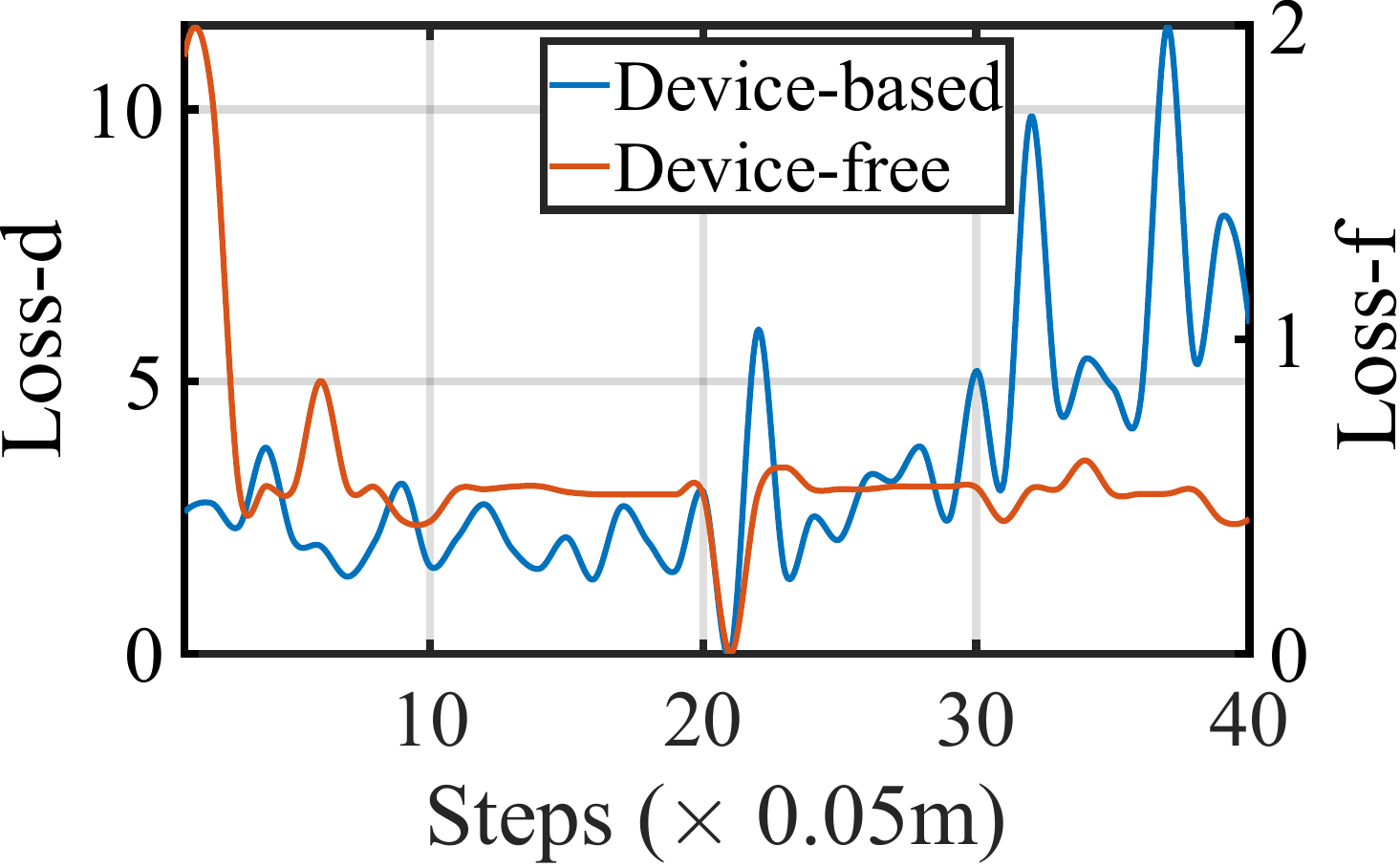}
                    
			\end{minipage}
			\label{fig:moti-loss}
		}
		\caption{Optimization-unfriendly loss landscape of RT.}
		\label{fig:moti-spr-gra}	
\end{figure}

The optimization problem in forward RT localization is non-convex, significantly limiting the effectiveness of gradient-based methods. To demonstrate the difficulty of optimization, we conduct experiments in both device-free and device-based scenarios using the RT simulator. The layout of the sensing scene is illustrated in Fig.~\ref{fig:moti-scene}. For device-free localization, we use a vase as the localization target. Initially, the vase's ground-truth location is at coordinates~(-1~\!m, 0~\!m). We relocate the vase to coordinates~(-1~\!m, 1~\!m) and move the vase along the $y$-axis to the coordinates~(-1~\!m, -1~\!m) with a step size of 0.05~\!m. Under the device-based setting, we also move the device along the direction vector (-2~\!m, -1~\!m) to the coordinates~(-2.5~\!m, -2~\!m) with a step size of 0.056~\!m, initializing the device's position at coordinates~(-0.5~\!m, -1~\!m) and the ground-truth position is at coordinates~(-1.5~\!m, -1.5~\!m). The CSI when the device is at the ground-truth location is recorded as a reference value. We take the mean squared error (MSE) between the reference CSI and the simulated CSI as the loss. The procedure of position change and the associated loss are depicted in Fig.~\ref{fig:moti-spr-gra}. Loss-d represents the loss in device-based settings, while loss-f represents the loss in device-free settings.

As illustrated in Fig.~\ref{fig:moti-loss}, as the device continues to move, the loss exhibits continuous fluctuations. Within 20 to 40 steps, the loss shows significant fluctuations with an upward trend. However, within 0 to 20 steps, the loss remains relatively stable and is dominated by local fluctuations. When using gradient-based approaches, it is difficult to provide effective gradients to navigate the device to the correct position. As for the device-free localization, we can observe that changes in object position exert a minor effect on the loss, in contrast to variations in device-based settings. This results in a nearly flat loss curve in certain regions when the object is not close to the ground-truth position. This leads to insufficient gradient in the region. The gradient-based optimizer is not able to position the vase correctly due to the plateau in the loss landscape. Consequently, both device-free and device-based localization face the challenge of sparse gradients and local minima. For the sake of accurate localization, it is imperative for us to improve the optimization properties of RT. 

\begin{figure*}[!h] %
    \centering %
    \includegraphics[width=0.95\textwidth]{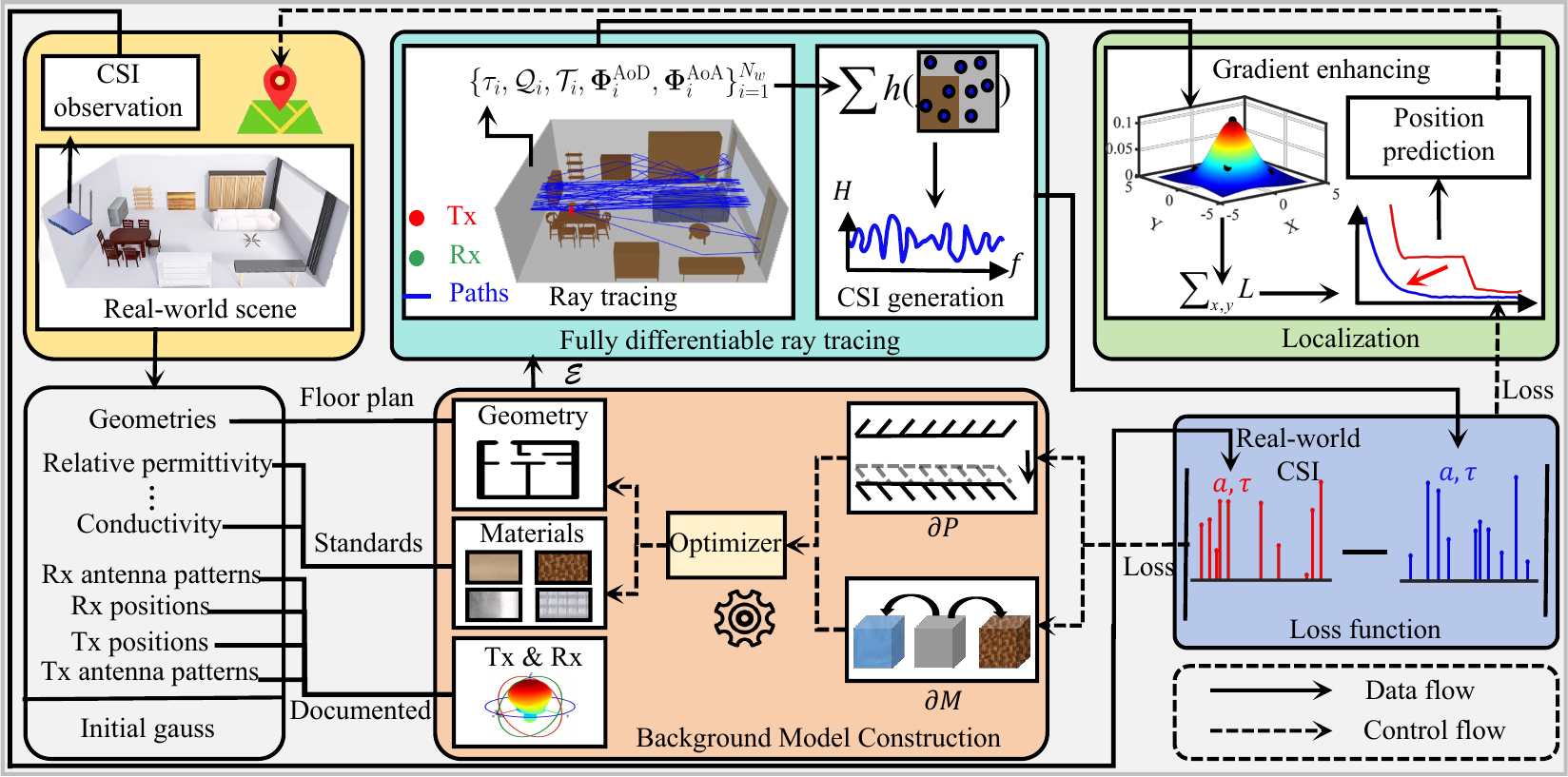} %
    \caption{The workflow of \name.} %
    \label{fig:workflow} %
    \vspace{-3ex}
\end{figure*}

\section{System Design} \label{sec:design}
In this section, we present \name's design. We first explain the problem formulation of \name. Subsequently, we explain the 3 steps of \name's workflow, i.e., fully differentiable RT, high-fidelity background model construction, and gradient-enhanced localization, as shown in Fig.~\ref{fig:workflow}. %

\subsection{Problem Formulation}
In this section, we explain the problem formulation of \name. At its core, \name solves an inverse problem of differentiable wireless ray-tracing that encompasses both scene parameter estimation and target localization. The scene is characterized by its EM environment parameters $\mathbf{I} = f(\bm{\mathcal{G}}, \bm{\mathcal{M}}, \mathbf{P}^\text{T}, \mathbf{P}^\text{R}, C^\text{T}, C^\text{R})$, which incorporate four fundamental components: geometric properties $\bm{\mathcal{G}}$, electromagnetic (EM) properties of materials $\bm{\mathcal{M}}$, transmitter and receiver positions $\mathbf{P}^\text{T}$ and $\mathbf{P}^\text{R}$, and their respective antenna patterns $C^\text{T}$ and $C^\text{R}$.

For a scene containing $N$ objects, each object $i$ is fully described by its position $\mathbf{P}_i$, shape $\mathbf{S}_i$, and material properties $\mathbf{M}_i$. The shape $\mathbf{S}_i$ is represented by a 3D triangle mesh comprising a set of vertices $\bm{\mathcal{V}}$ and a set of triangular faces $\bm{\mathcal{F}}$ that connect these vertices. Each triangular face specifies its three vertices from $\bm{\mathcal{V}}$, formally expressed as $\bm{\mathcal{F}}\subset \bm{\mathcal{V}} \times \bm{\mathcal{V}} \times \bm{\mathcal{V}}$~\cite{3d-mesh}. The complete scene model can thus be characterized by its geometry $\bm{\mathcal{G}} = \{\mathbf{P}_i, \mathbf{S}_i\}_{i = 1}^{N}$ and material EM properties $\bm{\mathcal{M}} = \{\mathbf{M}_i\}_{i=1}^{N}$. The target location $\mathbf{P}$ can represent either an object position $\mathbf{P}_i$, transmitter position $\mathbf{P}^\text{T}$, or receiver position $\mathbf{P}^\text{R}$, accommodating both device-free and device-based localization scenarios.

We approach this comprehensive inverse problem through a two-stage optimization framework. In the first stage, detailed in \S~\ref{ssec:calibration}, we estimate the background scene parameters $\mathbf{I} \setminus {\mathbf{P}}$, explicitly excluding the target location $\mathbf{P}$. This stage establishes a high-fidelity electromagnetic context essential for accurate localization. The second stage, presented in \S~\ref{sec:grad-enh}, leverages this high-fidelity background scene to perform precise estimation of the target location $\mathbf{P}$.

\subsection{Fully differentiable Wireless RT} \label{ssec:fully_diff}
In this section, we present a wireless RT method that is fully differentiable, thus serving as a foundation for inverse optimization.
\subsubsection{Wireless RT} \label{SBR}
\begin{figure}[H]
    \centering
    \includegraphics[width=0.48\textwidth]{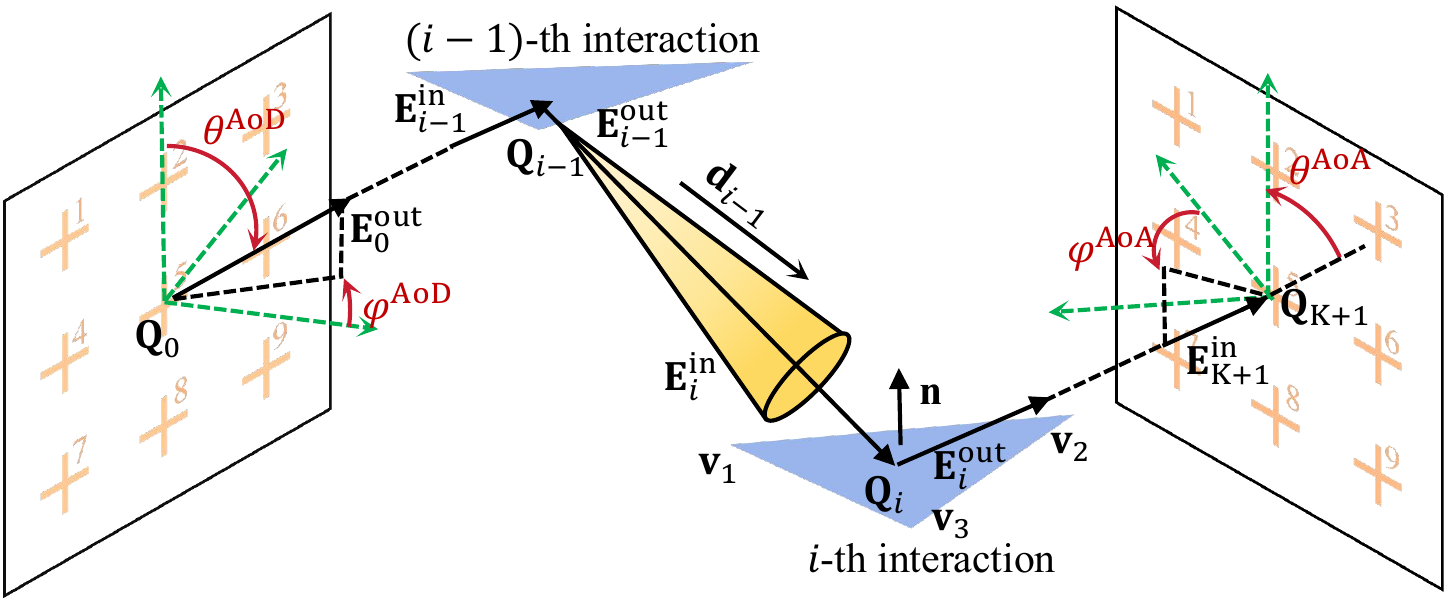}
    \caption{Illustration of the SBR process.}
    \label{fig:paths}
\end{figure}

As illustrated in Fig.~\ref{fig:workflow}, RT takes the scene model $\mathbf{I}$ as input and outputs the propagation paths between the transceiver antenna pairs, serving as a core enabler for wireless simulations. In this paper, our simulation is conducted using the shooting and bouncing ray (SBR) method, as it is deemed a more computationally efficient RT approach~\cite{SBR}. The basic concept behind the SBR is to launch multiple rays from Tx's antenna, recursively track the propagation paths according to Snell's law, and find the paths that can reach the Rx's antenna within a given depth, i.e., the maximum number of interactions $t_\text{max}$. Fig.~\ref{fig:paths} illustrates the process of SBR for a specific ray and the output propagation path. 

Each ray emitted from the Tx is defined by its origin $\mathbf{Q}_0$ and an initial normalized orientation vector $\mathbf{d}_0$. 
The emitted ray marches a distance $t_0$ along its initial direction $\mathbf{d}_0$, interacting with the triangular primitive of an object in the scene model. The ray geometry can be represented as $\mathbf{O}(t) = \mathbf{Q}_0 + t_0\mathbf{d}_0$. After interacting with the triangular face, the ray gains a new origin $\mathbf{Q}_1$ and orientation vector $\mathbf{d}_1$.
Iteratively, the origin of the ray incident on the $i$-th triangular face is $\mathbf{Q}_{i-1}$, and its direction is $\mathbf{d}_{i-1}$ after $i-1$ interactions with the scene model. 
Let the $i$-th face be represented by three vertices $\mathbf{v}_1, \mathbf{v}_2, \mathbf{v}_3$. The normal vector $\mathbf{n}$ of the face is a unit vector that can be calculated by $\overrightarrow{\mathbf{v}_1\mathbf{v}_2} \times \overrightarrow{\mathbf{v}_2\mathbf{v}_3}$. The new origin $\mathbf{Q}_i$ is essentially the intersection point of the ray and the triangle, which should satisfy the ray geometry equation:
\begin{equation}
    \mathbf{Q}_{i} = \mathbf{Q}_{i-1} + t_{i-1}\mathbf{d}_{i-1},
    \label{ray-eqn}
\end{equation}
where $t_{i-1}$ is the distance from $\mathbf{Q}_{i-1}$ to $\mathbf{Q}_i$ along the ray orientation vector $\mathbf{d}_{i-1}$. Additionally, the intersection point $\mathbf{Q}_i$ is also on the $i$-th face, therefore
\begin{equation}
    \mathbf{Q}_i = (1-m_1-m_2)\mathbf{v}_3 + m_1\mathbf{v}_1 + m_2\mathbf{v}_2,
    \label{plane-eqn}
\end{equation}
where $0\leq m_1, m_2 \leq 1$ are the weight for vertices. 
When $t_{i-1} > 0, m_1 > 0, m_2 > 0$  is satisfied, the intersection point $\mathbf{Q}_i$ is within the triangle surface and can be obtained with the Eq. \ref{ray-eqn} and \ref{plane-eqn}.
The direction of the outgoing ray $\mathbf{d}_i$ is determined by the physical law of wave propagation. In this paper, we focus on the simulation of reflection paths. The reflected direction of the outgoing ray is as follows:
\begin{equation}
    \mathbf{d}_i = \mathbf{d}_{i-1} - 2(\mathbf{d}_{i-1}^{\text{T}}\mathbf{n})\mathbf{n}.
\end{equation}
Finally, the ray is captured by the antennas of the Rx. 

A wireless ray represents more than just a geometric path, it also carries EM energy as part of the propagating EM field. The EM field $\mathbf{E}$ can be represented in the form of two different orthogonal polarization components $\mathbf{E}_s$ and $\mathbf{E}_p$, i,e.,
\begin{equation}
    \mathbf{E} = \mathbf{E}_s \mathbf{e}_s + \mathbf{E}_p \mathbf{e}_p,
\end{equation}
where $\mathbf{e}_s$ and $\mathbf{e}_p$ are unit vectors denoting the two different orthogonal polarization directions.
The radiation EM field from Tx denoted as $\mathbf{E}_0$ depends on the antenna pattern of the Tx, i.e., $\mathbf{E}_0^{\text{out}} = C^\text{T}$.
After traveling a distance $t_i$, the phase of the EM field changes to $e^{-j2\pi f t_i}\mathbf{E}_i$, where $f$ is the frequency of the EM wave. The total EM energy of the ray is constant in free space, but the energy density decreases when the ray travels, as shown by the circular ray tube in Fig.~\ref{fig:paths}. 
Then, the EM field is distorted by the interactions with the $i$-th triangle that separates two materials, e.g., air and material of the triangular face. The distorted field $\mathbf{E}_i^{\text{out}}$ can be written as $\mathbf{E}_i^{\text{out}} = \mathbf{T}_i(\mathbf{E}_i^\text{in}) = F_i(\mathbf{D}_i\mathbf{E}_i^{\text{in}})$, 
where $\mathbf{D}_i$ is a basis transformation matrix that converts the incident field into the local coordinate system of the $i$-th triangle, i.e., 
\begin{equation}
    \mathbf{D}_i = 
    \begin{bmatrix}
     \mathbf{e}_{i,\perp}^\text{T}\mathbf{e}_{i,s} & \mathbf{e}_{i,\perp}^\text{T}\mathbf{e}_{i,p}\\
    \mathbf{e}_{i,\parallel}^\text{T}\mathbf{e}_{i,s} & \mathbf{e}_{i,\parallel}^\text{T}\mathbf{e}_{i,p}
\end{bmatrix}.
\end{equation}
Here, $\mathbf{e}_{i,s}$ and $\mathbf{e}_{i,p}$ are the two different orthogonal polarization directions of $\mathbf{E}_i^\text{in}$. $\mathbf{e}_{i,\perp} = \frac{\mathbf{d}_{i-1} \times \mathbf{n}}{\| \mathbf{d}_{i-1} \times \mathbf{n}\|}$ and $\mathbf{e}_{i,\parallel} = \mathbf{e}_{i,\perp} \times \mathbf{n}$ are the two different orthogonal polarization directions of transformed $\mathbf{E}_i^\text{in}$. $\mathbf{e}_{i,\perp}^\text{T}$ and $\mathbf{e}_{i,\parallel}^\text{T}$ are the transpose of $\mathbf{e}_{i,\perp}$ and $\mathbf{e}_{i,\parallel}$, respectively. $F_i$ describes the relationship between the incoming and outgoing fields after interacting with the $i$-th triangle, which can be written as:
\begin{equation}
    F_i(\mathbf{E}) = \begin{bmatrix}
        r_\perp & 0 \\
        0 & r_\parallel
    \end{bmatrix} \mathbf{E} A^r(\mathbf{Q}_{j-1}, \mathbf{Q}_j) e^{-j2\pi f t_j},
\end{equation}
where $A^r(\mathbf{Q}_{j-1}, \mathbf{Q}_j)$ is the spreading factor that depends on the shape of the wavefront~\cite{ray-tech} and $r_\perp$ and $r_\parallel$ are the Fresnel reflection coefficients~\cite{itu}:
\begin{equation}
    \begin{aligned}
        r_\perp= \frac{\cos(\theta) - \sqrt{\eta - \sin^2(\theta)}}{\cos(\theta) + \sqrt{\eta - \sin^2(\theta)}}, \quad r_\parallel= \frac{\eta \cos(\theta) - \sqrt{\eta - \sin^2(\theta)}}{\eta \cos(\theta) + \sqrt{\eta - \sin^2(\theta)}},
    \end{aligned}
\end{equation}
where $\cos(\theta) = -\mathbf{d}_{i-1} \times \mathbf{n}$ and the $\eta = \epsilon_r - j\frac{\sigma}{2\pi f\epsilon_0}$ is the complex relative permittivity that is associated with the relative permittivity $\epsilon_r$ and the conductivity $\sigma$ of the material~\cite{itu}. After undergoing $K$ interactions, the EM field is ultimately received by the Rx. 
The transformation processes of the EM field can be represented by a single matrix $\bm{\mathcal{T}}$~\cite{learnable, sionna-rt}. The ultimate energy received $\mathbf{E}_{K+1}$ can be written as:
\begin{equation}
   \begin{aligned}
        \mathbf{E}_{K+1}^\text{in} = \mathbf{T}_K\left(...\left(\mathbf{T}_1\left(\mathbf{T}_0(\mathbf{E}_0^\text{out})\right)\right)\right) = \bm{\mathcal{T}}\left(C^\text{T}(\mathbf{\Phi}^\text{AoD})\right)e^{-j2\pi f\tau},
\end{aligned} 
\end{equation}
where the $\mathbf{E}_0^\text{out}$ is the initial energy of the ray when emitting from the Tx. The $\mathbf{\Phi}^\text{AoD} = (\theta^\text{AoD},\phi^\text{AoD})$ is the angles of departure (AoD) from the Tx antenna, including the elevation $\theta^\text{AoD}$ and azimuth $\phi^\text{AoD}$ angles in the Tx's coordinate system, as shown in Fig.~\ref{fig:paths}. The $e^{-j2\pi f\tau}$ depicts the phase shift of the path with delay $\tau$, where $\tau = \frac{1}{c}(t_0+t_1+...+t_K)$. The $c$ is the vacuum speed of light. $\mathbf{T}_i$ is a function that depends on the interaction type (e.g., reflection), the geometric properties of the interaction (including the hit point $\mathbf{Q}_i$ and the direction $\mathbf{d}_i$), and the EM characteristics of the material involved, for $i = \{0,1, ..., K\}$. The $\mathbf{Q}_0 = \mathbf{P}^\text{T}$ and $\mathbf{Q}_{K+1} = \mathbf{P}^\text{R}$ denote the position of the Tx and Rx, respectively.

With the SBR method, we can obtain the parameters of all the traced $N_w$ propagation paths given by $\{\tau_i, \bm{\mathcal{Q}}_i, \bm{\mathcal{T}}_i, \bm{\Phi}_i^\text{AoD}, \bm{\Phi}_i^\text{AoA}\}_{i=1}^{N_w}$, where each path $w_i$ for $i\in\{1, 2, ..., N_w\}$ is described by a set of intersection points $\bm{\mathcal{Q}}_i = \{\mathbf{Q}_0, \mathbf{Q}_1, ...., \mathbf{Q}_{K+1}\}$, a delay $\tau_i$, a pair of angles of arrival $\mathbf{\Phi}_i^\text{AoD}$, and a pair of angles of arrival $\mathbf{\Phi}_i^\text{AoA}$. The $\mathbf{\Phi}_i^\text{AoA} = (\theta_i^\text{AoA},\phi_i^\text{AoA})$ is the angles of the $i$-th path captured by Rx, encompassing the elevation and azimuth angles, denoted by $\theta_i^\text{AoA}$ and $\phi_i^\text{AoA}$, respectively.

\subsubsection{CSI Generation} \label{csi-gen}
In the previous section, we investigate how paths propagate from the Tx's antennas to the Rx's antennas. Ideally, we should integrate all plausible paths between antenna of the Tx and Rx pair to obtain an accurate simulation of the CSI. Formally, we can model the CSI with frequency $f$ as:
\begin{equation}
    h(\mathbf{I}, f) = \int_{\mathbf{\Omega}} \mathcal{H}(\mathbf{I}, f, w) dw,
    \label{int-csi}
\end{equation}
which is an integral of the function $\mathcal{H}$ that depends on the scene model $\mathcal{E}$ configurations, over all possible paths $w\in\Omega$. However, there is no closed-form solution to Eq. \ref{int-csi}. Therefore, Monte Carlo (MC) methods are typically used to estimate the integral by randomly sampling $N_w$ paths across $\mathbf{\Omega}$:
\begin{equation}
    \hat{h}(\mathbf{I}, f) = \frac{1}{N_w}\sum_{i=1}^{N_w} \mathcal{H}(\mathbf{I}, f, w_i),
\end{equation}
where $\hat{h}$ converges to $h$ as $N_w \rightarrow \infty$. 

Let $h_{n_t, n_r}$ be the CSI between the $n_t$-th antenna and the $n_r$-th antenna at frequency $f$. Given the parameters of the $N_w$ paths at frequency $f$ between a pair of antennas, each associated with the Tx and Rx, $h_{n_t, n_r}$ can be defined as
\begin{equation}
    h_{n_t, n_r}(\mathbf{I}, f) = \sum_{i = 1}^{N_w} C^\text{R}(\mathbf{\Phi}_i^\text{AoA}) \bm{\mathcal{T}}_i\left(C^\text{T}(\mathbf{\Phi}_i^\text{AoD})\right)e^{-j2\pi f\tau_i}.
    \label{eqn:csi}
\end{equation}
Wireless systems typically feature multiple antennas and use Orthogonal Frequency Division Multiplexing (OFDM) to divide their bandwidth into several orthogonal subcarriers. So, the CSI obtained from the Tx/Rx is a complex-valued matrix of size $N_t\times N_r\times N_s$. Here, $N_t$ and $N_r$ are the number of Tx and Rx antennas; $N_s$ is the number of subcarriers. Let $f_c$ and $\Delta f$ be the central frequency and the frequency spacing between the two adjacent subcarriers, respectively. The estimated CSI matrix $\mathbf{H}$ is:
\begin{equation}
    \mathbf{H} = \left[h_{n_t, n_r}(\mathbf{I}, f_0),..., h_{n_t, n_r}(\mathbf{I}, f_j), ..., h_{n_t, n_r}(\mathbf{I}, f_{N_s})\right]_{N_t\times N_r},
\end{equation}
where $f_j = f_c + j\frac{N_s}{2}\Delta f$ is the frequency of the $j$-th subcarrier. 

\subsubsection{Trainable Scene Parameters}
After obtaining the CSI with the forward model, we can define a loss function $L$ to optimize the scene parameters $\mathbf{I}$ with a gradient-based approach. 
The key to the optimization lies in correctly computing the gradient of the loss function $L$ w.r.t. the scene parameters $\mathbf{I}$. 
We can consider the forward model described in \S~\ref{SBR} and \S~\ref{csi-gen} as a complex function affected by scene parameters $\mathbf{I}$ (including both transceiver and object location for device-free and device-based localization, respectively) and implement it using a machine-learning framework such as TensorFlow and PyTorch. To obtain the gradients, we can make the scene parameters trainable and leverage the automatic differentiation capabilities of the ML framework to compute the gradients.
According to the chain derivation rule, the gradient of the scene parameters can be written as:
\begin{equation}
    \frac{\partial L}{\partial \mathbf{I}} = \frac{\partial L}{\partial \mathbf{H}} \frac{\partial \mathbf{H}}{\partial \mathbf{I}}.
\end{equation}
A differentiable loss function w.r.t. $\mathbf{H}$ can be crafted manually. As for the $\frac{\partial \mathbf{H}}{\partial \mathbf{I}}$, it is obtainable in general cases, given that both the MC and the transfer function $\bm{\mathcal{T}}$ are differentiable w.r.t. $\mathbf{I}$~\cite{sionna-rt}. However, the MC sampling process is not continuous in the case of visibility changes as the sampled paths vary. 
Occlusion by target objects during localization can cause sudden visibility changes, making it difficult to differentiate between different localization states. To overcome the problem, we employ a smoothing technique to avoid abrupt visibility changes~\cite{smoothing-1, smoothing-2}. Specifically, we replace discrete decisions, such as testing the validity of paths, with soft functions, mitigating the impact of changes in visibility.

\subsection{High-fidelity Background Model Construction}\label{ssec:calibration}
In this section, we elaborate on how the high-fidelity background model is constructed, especially, how the scene geometry and EM material properties are obtained.
\subsubsection{Input Data and Inductive Bias}
To achieve high-precision background modelling, we can leverage a set of CSI measured from the ground-truth scene as the reference to calibrate these scene parameters. We assume the available CSI dataset is $\bm{\mathcal{D}} = \{\mathbf{P}^r_i, \mathbf{H}_i\}_i^{N_d}$, where $\mathbf{H}_i$ is the collected CSI when the Rx is positioned at coordinate $\mathbf{P}^r_i=(x_i, y_i, z_i)$ and $N_d$ is the number of the measured CSI. We can extract features like the amplitude and phase from the complex-valued CSI to form a specific loss function $L$. Similar to training DL models, gradient-based optimization approaches can be used to minimize the loss function $L$ on dataset $\bm{\mathcal{D}}$, thereby obtaining more accurate scene parameters $\mathbf{I}^*$, i.e.,
\begin{equation}
    \mathbf{I}^* = \argmin_{\mathbf{I}} L(\mathbf{I}|\bm{\mathcal{D}}).
\end{equation}

Employing a data-driven approach to calibrate these scene parameters appears straightforward. However, it remains challenging to achieve good performance even with stable real-world measurements. This is because a single path can interact with various objects before being captured by the Rx. Consequently, the position deviations of these objects and variations in the involved materials can influence each other during model construction, potentially leading to overfitting with local minima~\cite{material-rendering}. Therefore, we need to introduce inductive biases into the model construction process, i.e., to make several prior assumptions about the optimization to prompt plausible results. First, we set the related characteristics to materials based on common sense and choose the corresponding values from the ITU-R P.2040-2~\cite{itu} standards rather than random initialization. Second, positional deviations of objects within several centimeters primarily cause changes in the path length, whereas material variations are independent of the path length~\cite{phase-calibration, csi-ratio}.

\subsubsection{Loss Functions for Background Model Construction}
There are many different types of scene parameters, including object positions and material characteristics (the relative permittivity and the conductivity).
All of these scene parameters significantly affect the generation of CSI, making it impractical to optimize all these parameters simultaneously with a unified loss function. To address the challenge, we use different loss functions to calibrate the object position and EM properties of the materials involved in an alternate manner. Since the path length is independent of the material, we focus more on length-related changes to optimize object position, while concentrating on amplitude variations to optimize material characteristics. 
However, random noise is inevitably introduced in real-world measurements due to the lack of time synchronization between the Tx and Rx, as well as environment dynamics. There is a time-varying random phase offset in these measurements. 
We use the ratio of the CSI between two adjacent antennas to mitigate noise interference, as the time-varying offsets are identical across different antennas of the same Tx/Rx~\cite{csi-ratio}. The CSI ratio is as follows:
\begin{equation}
    \Delta h_{1,2}(\mathbf{I},f)= \frac{\Tilde{h}_1(\mathbf{I},f)}{\Tilde{h}_2(\mathbf{I},f)} = \frac{e^{-j\phi_\text{off}}h_1(\mathbf{I},f)}{e^{-j\phi_\text{off}}h_2(\mathbf{I},f)} = \frac{h_1(\mathbf{I},f)}{h_2(\mathbf{I},f)},
\end{equation}
 where $h(\mathbf{I}, f)$ is the true CSI and $e^{-j\phi_{\text{off}}}$ is the phase offset. 
After obtaining the CSI ratio, we apply the Savitzky-Golay filter to smooth it. 

Therefore, the designed loss function to optimize the object position is as follows:
\begin{equation}
    L_p = \frac{\sum_i^{N_\text{tot}} \tau_i - \sum_i^{N_\text{tot}} \hat{\tau}_i}{\sum_i^{N_\text{tot}} \tau_i + \sum_i^{N_\text{tot}} \hat{\tau}_i},
\end{equation}
where $N_\text{tot} = N_t \times N_r\times N_s$ is the dimension of the CSI matrix. $\tau_i$ is obtained from the real-world measurements, while $\hat{\tau}_i$ is obtained from simulated CSI. The function 
\begin{equation}
    l(x, y) = \frac{x - y}{x + y} \in [0, 1]
\end{equation}
is the symmetric mean absolute percentage error (SMAPE) that is chosen to normalize the measurements taken at different positions in the scene ~\cite{sionna-rt}. As for the loss function for EM properties of materials, we design as follows:
\begin{equation}
    L_m = \frac{\sum_i^{N_\text{tot}^{\prime}} a_i - \sum_i^{N_\text{tot}^{\prime}} \hat{a}_i}{\sum_i^{N_\text{tot}^{\prime}} a_i + \sum_i^{N_\text{tot}^{\prime}} \hat{a}_i},
\end{equation}
where $a_i$ is the amplitude of the CSI ratio obtained from the real world, while $\hat{a}_i$ is the amplitude of the CSI ratio obtained from simulation. $N_\text{tot}^{\prime} = (N_t - 1) \times N_r\times N_s$ is the dimension of the CSI ratio matrix.

\subsection{Gradient-enhanced Localization}\label{sec:grad-enh}
In this section, we dive into the proposed wireless indoor localization approach with gradient-based optimization. 
\footnote{While both background model construction (\S~\ref{ssec:calibration}) and localization employ gradient-based optimization, their operations are signicantly different: the former requires fine adjustments within wavelength-scale distances, whereas the latter must determine positions across several meters. This larger scale makes localization particularly vulnerable to local minima in the loss landscape (\S~\ref{sec:plateaus}).} 
To tackle the non-convex optimization, we employ a smoothing technique in the loss function and develop an efficient method to address the localization problem based on the inverse process of ray-tracing. %

\subsubsection{Loss Function for Localization}
To achieve accurate localization with the gradient-based approach, the key first step is to craft a loss function $L$. In our paper, we define the loss function $L$ that incorporates both the amplitude $\{a_i\}_{i=1}^{N_\text{tot}^{\prime}}$ of the CSI ratio and delay $\{\tau\}_{i=1}^{N_\text{tot}}$ of the CSI, i.e.,
\begin{equation}
       L(\mathbf{I}, f) = \mathrm{MSE}(a_i, \hat{a}_i) + \gamma_1 \mathrm{MSE}(\tau_i - \hat{\tau}_i) + \gamma_2 (x^2+y^2), 
\end{equation}
where $\mathrm{MSE}$ represents the mean squared error, a widely used function to gauge the error between predicted and ground-truth values. The term $(x^2+y^2)$ serves as a regularization term for the current position $(x, y)$, preventing overfitting to the position and aiding in escaping local optima. The $\gamma_1$ and $\gamma_2$ are scaling factors that ensure the numerical comparability of these loss terms, preventing the model from overemphasizing amplitude changes. 

The reason behind using a loss function based on the SMAPE in background model construction and the MSE in localization lies in their distinct roles within the process. The SMAPE primarily serves to normalize CSI from multiple measurement points, ensuring that all examples are weighted equally~\cite{sionna-rt}. On the other hand, the MSE offers superior convexity, which is more conducive to optimization in object localization. 
Hence, we adopt two distinct functions, SMAPE and MSE, for the background model construction and localization tasks, respectively.

\subsubsection{Coarse-to-Fine Gaussian Smoothing}
 Using the gradient descent method with the loss function $L$ for localization is straightforward. It can be formulated as follows:
\begin{equation}
    \mathbf{P}^{\prime}_{x,y} = \mathbf{P}_{x,y} - \epsilon \frac{\partial L}{\partial \mathbf{P}_{x,y}},
\end{equation}
where $\mathbf{P}_{x,y}$ represents a 3D coordinate within the region where the object can be placed at arbitrary positions on the 2D plane $\mathbf{X} \times \mathbf{Y}$, while its height $z$ remains constant. $\epsilon$ is the learning rate for gradient update.
However, it might yield incorrect localization results, especially in device-free settings, as shown in \S~\ref{sec:plateaus}. To improve the optimization properties, we employ a smoothing transformation to the loss function to facilitate the search for the exact object position. Theoretically, we convolve the loss function $L$ with a smoothing kernel $g$ over the object position, which can be defined as
\begin{equation}
    L_{\bm{\sigma}}(\mathbf{I}, f) = \int_{\mathbf{X}\times \mathbf{Y}} g(x, y) L(\mathbf{I}-\mathbf{P}_i, \mathbf{P}_{x,y}, f)dxdy,
\end{equation}
where $\mathbf{P}_i$ is the initial position of the target object $i$. The kernel $g(x,y)$ is a 2D Gaussian kernel function $\mathcal{N}_\sigma(x, y) = \frac{1}{2\pi\sigma^2}\exp\left(\frac{-x^2 + y^2}{2\sigma^2}\right)$ that is a continuous, symmetric, and non-negative function. Here, the 2D Gaussian kernel is used as we consider object localization on a 2D plane while extending to 3D within our framework is also straightforward. The effectiveness of the method stems from the Gaussian kernel assigning weights to loss values at different positions of the object. The region near the object position $(x, y)$ is given higher weights, while the control of the variance allows it to modulate the weights of more distant positions. When calculating the loss, local information is maintained while still accounting for the influence of other positions. Consequently, the object can more easily escape from the plateau and local minima and locate the global optimal position.
\begin{figure}[t]
	\captionsetup[subfigure]{justification=centering}
		\centering
		\subfloat[Enhanced loss under the device-free setting.]{
        \hspace{-2ex}
		  \begin{minipage}[b]{0.49\linewidth}
		        \centering
			    \includegraphics[width = \textwidth]{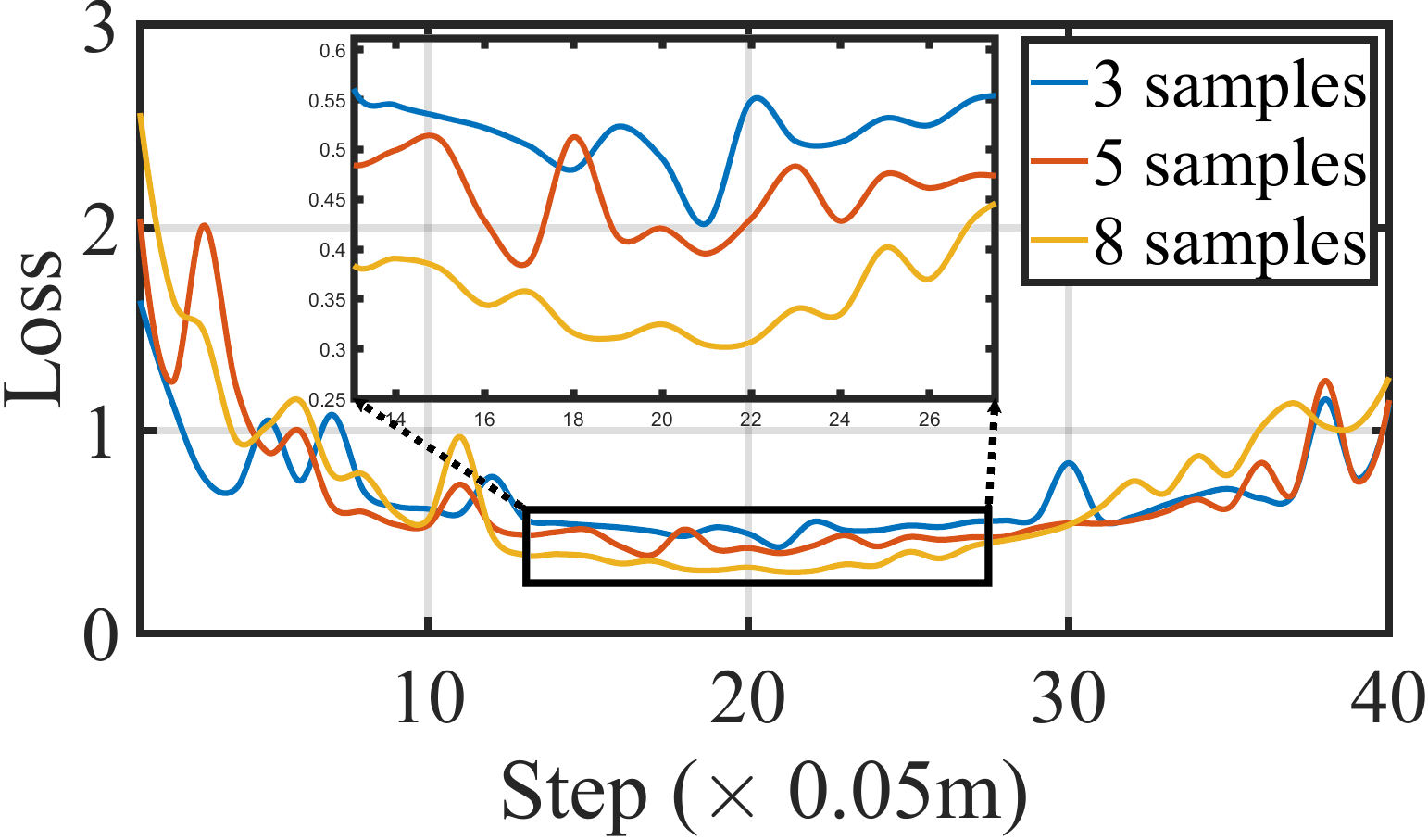}                   
			\end{minipage}
			\label{subfig:free-loss}
		}
		\subfloat[Enhanced loss under the device-based setting.]{
		    \begin{minipage}[b]{0.49\linewidth}
		        \centering
			    \includegraphics[width = \textwidth]{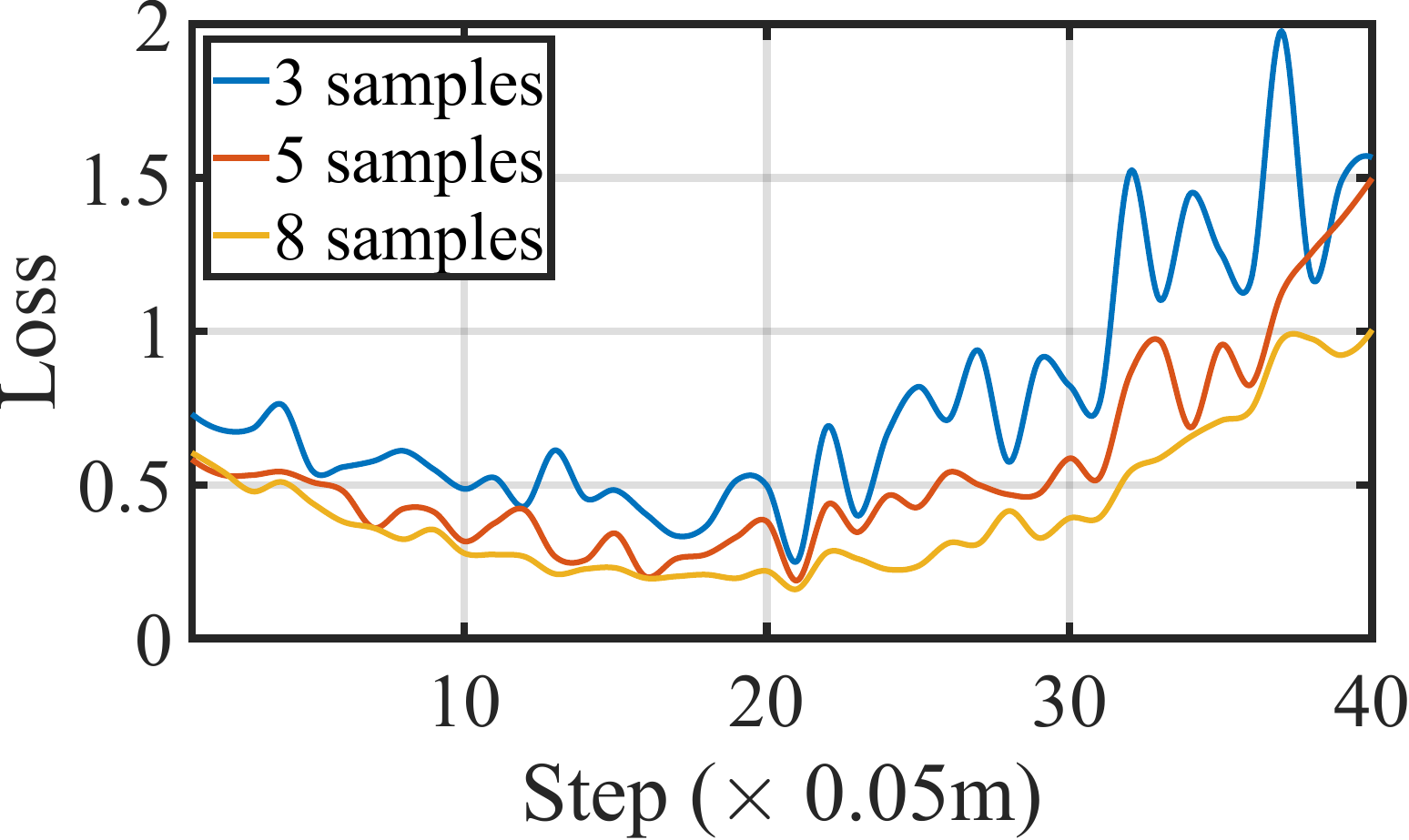}
                    
			\end{minipage}
			\label{subfig:based-loss}
		}
		\caption{The enhanced loss landscape with a different number of sampled positions under device-free and device-based settings.}
		\label{fig:smoothing}	
\end{figure}

Let $\bm{\mathbb{U}} = \mathbf{X} \times \mathbf{Y} \in \bm{\mathbb{R}}^2$ and $\mathbf{u} = (x, y) \in \mathbb{U}$. To estimate the enhanced loss $L_\sigma$, we can also use MC sampling across the $\bm{\mathbb{U}}$:
\begin{equation}
    \hat{L}_\sigma(\mathbf{I}, f)  = \frac{1}{N_p}\sum_{j=1}^{N_p} g(\mathbf{u}_j)L(\mathbf{I}-\mathbf{P}_i, \mathbf{P}_{\mathbf{u}_j}, f),
\end{equation}
where the $N_p$ represents the number of sampled 2D positions in the region $\mathbb{U}$, specifically the whole space of the indoor scene in our literature. 

However, sampling uniformly across $\mathbb{U}$ results in samples of greater importance that produce lower loss value being assigned unreasonable weights, leading to high-variance gradient predictions and ultimately slowing the convergence of optimization~\cite{BMR}. Therefore, we introduce a bias $B = \exp\left(-\alpha L(\mathbf{I}, f)\right)$ to assign greater weights to these key samples, accelerating convergence. The process can be formulated as
\begin{equation}
    \hat{L}_\sigma(\mathbf{I}, f)  = \frac{1}{N_p}\sum_{j=1}^{N_p} e^{-\alpha L_{\mathbf{u}_j}} g(\mathbf{u}_j)L(\mathbf{I}-\mathbf{P}_i, \mathbf{P}_{\mathbf{u}_j}, f).
\end{equation}
The bias $B$ treats the value of $L$ as a constant input, and it does not participate in the gradient computation. Thus, $B$ can be seen as a constant weight that shifts the input $\mathbf{u}$ towards low values of the function $L$. 
The gradient of the enhanced loss is actually a weighted sum of the true gradient in a neighborhood of the object position, i.e.,
\begin{equation}
    \frac{\partial L}{\partial \mathbf{u}} = \frac{1}{N_p}\sum_{j=1}^{N_p}B_j\frac{\partial}{\partial \mathbf{u}_j}g(\mathbf{u}_j)L_{\mathbf{u}_j}.
\end{equation}

Another key factor is the variance $\sigma$ that controls how far our samples will be spread out from the current position.
Initially, we need extensive exploration across the entire solution space $\mathbb{D}$, necessitating a large $\sigma$. However, as the optimization progresses, it becomes essential to gradually decrease the $\sigma$ close to $0$ to make more effective use of the better solution already explored so far. Hence, we decay the variance using a weighted linear strategy, i.e., $\sigma_{t+1} = \sigma_0 - \beta\frac{t}{T}(\sigma_0-\sigma_m)$. The $T$ is the total number of iterations, while $t$ represents the current iteration step.
The $\beta$ is a factor to control the decay speed of the $\sigma$. Notably, $\beta$ is the function with a value ranging from -2 to 2, formally defined as follows:
\begin{equation}
    \beta = 2 * \frac{1}{1+e^{\text{softmax}(B_c)}},
\end{equation}
where $\text{softmax}(B_c) = \frac{e^{B_c}}{\sum_{j=1}^{N_p}e^{B_j}}$, and $B_c = \exp(-\alpha L(I-P_i, P_i, f))$ is the value of the bias term at the current position. By adjusting the weights, the convergence of $\sigma$ can strike a more effective balance between exploration and exploitation, leading to faster convergence.

To demonstrate the effectiveness of the enhanced loss, we obtain it with different numbers of sampling positions, specifically 3, 5, and 8 samples, under the configurations described in \S~\ref{sec:plateaus}. The results are illustrated in Fig.~\ref{fig:smoothing}. We can observe that the enhanced loss curves are smoother, which facilitates gradient-based optimization. Additionally, as the number of samples increases, the loss landscape becomes even smoother, and the overall loss values decrease due to the bias term of the design towards smaller loss values.

\section{Implementation}\label{sec:impl}
For the scene model construction, we employ the floor plans to obtain a coarse-grained scene model and then import it into Blender 3.6 LTS~\cite{Blender3.6} to render a more authentic indoor environment for our experiments. Then, the scene model is exported from Blender in XML file format using the Mitsuba-Blender add-on~\cite{addon} to facilitate path propagation modeling in Mitsuba 3~\cite{Mitsuba3}. We implement \name using Python 3.8 and TensorFlow 2.13. As for the $\gamma$ in the loss function of localization, we set it to 2 and 0.05 for the device-free and device-based settings, respectively. We sample our Gaussian kernel with zero mean. The covariance of the kernel is set to 0.1 and 1 for the device-free and device-based settings, respectively. Both the minimum covariance $\sigma_m$ for the two settings is 0.01. 
The covariance $\alpha$ in the bias term is set to 1 and 0.2 for the device-free and device-based settings, respectively. 
The number of sampled 2D positions $N_p$ is set to 5. We use Root Mean Square Propagation (RMSprop) as our optimizer for background model construction and target localization. The learning rate and momentum are set to 0.03 and 0.6, respectively. In the localization configurations, the object to be localized is randomly initialized within a 2D space of the indoor environment, and the number of iterations $T$ for gradient descent is set to 100. 

\section{Evaluation}\label{sec:evaluation}
In this section, we first introduce the experiment setup and detail the selected baselines. To evaluate the performance of \name, we test it under varying experiment settings. In particular, we evaluate the performance of the proposed \name from three aspects: i) we compare \name with six baselines to demonstrate its superiority; ii) we test \name under various environment settings to show its robustness; iii) ablation study is conducted to show the necessity of key module designs.
\subsection{Experiment Setup}
\begin{figure}[b]
    \setlength\abovecaptionskip{6pt}
    \vspace{-2ex}
	\captionsetup[subfigure]{justification=centering}
		\centering
		\subfloat[The meeting room.]{
            \hspace{-2ex}
		  \begin{minipage}[b]{0.49\linewidth}
		        \centering
			    \includegraphics[width = \textwidth, height = 0.59\textwidth]{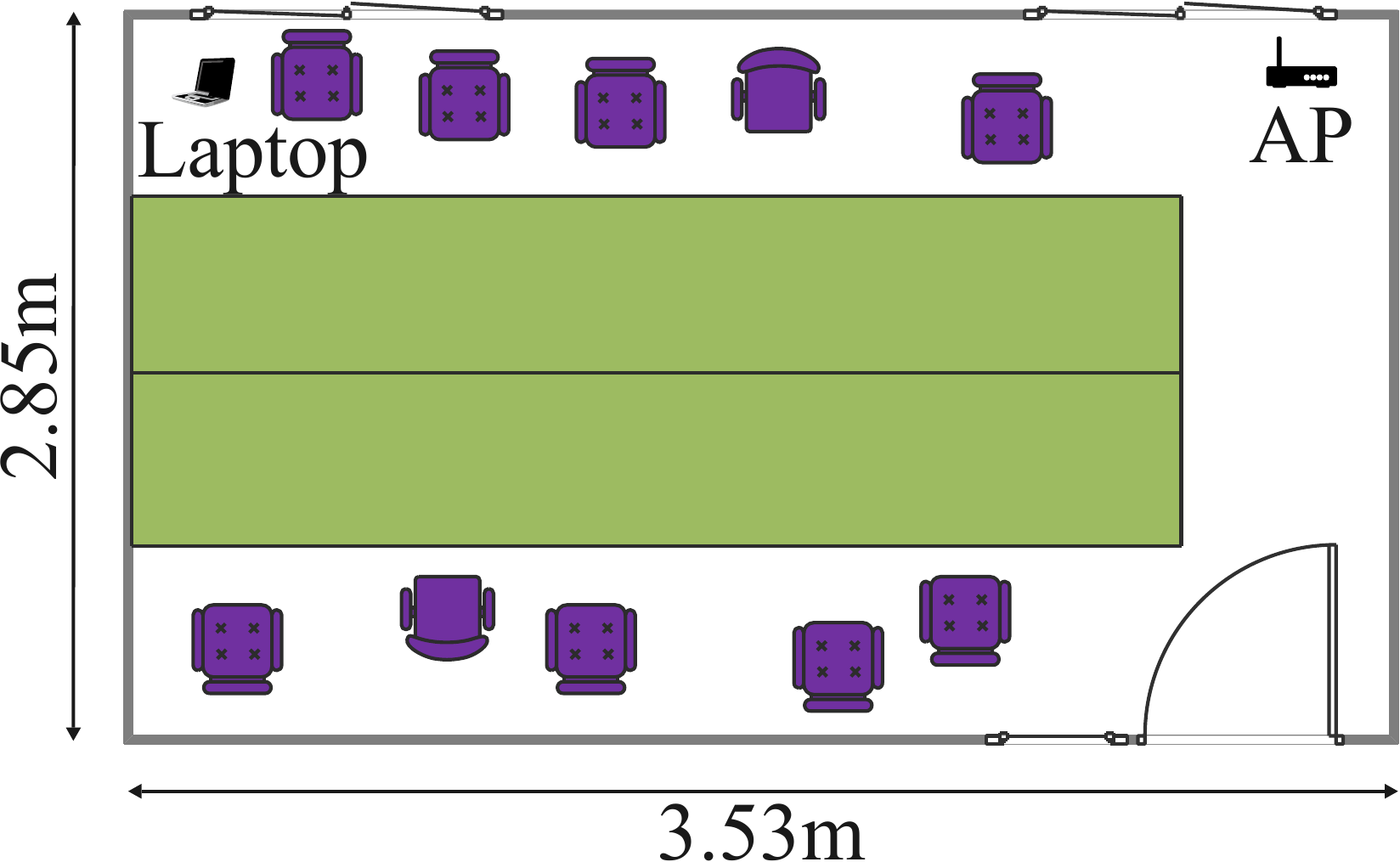}                   
			\end{minipage}
			\label{subfig:conference}
		}%
		\subfloat[The laboratory.]{
		    \begin{minipage}[b]{0.49\linewidth}
		        \centering
			    \includegraphics[width = \textwidth, height = 0.59\textwidth]{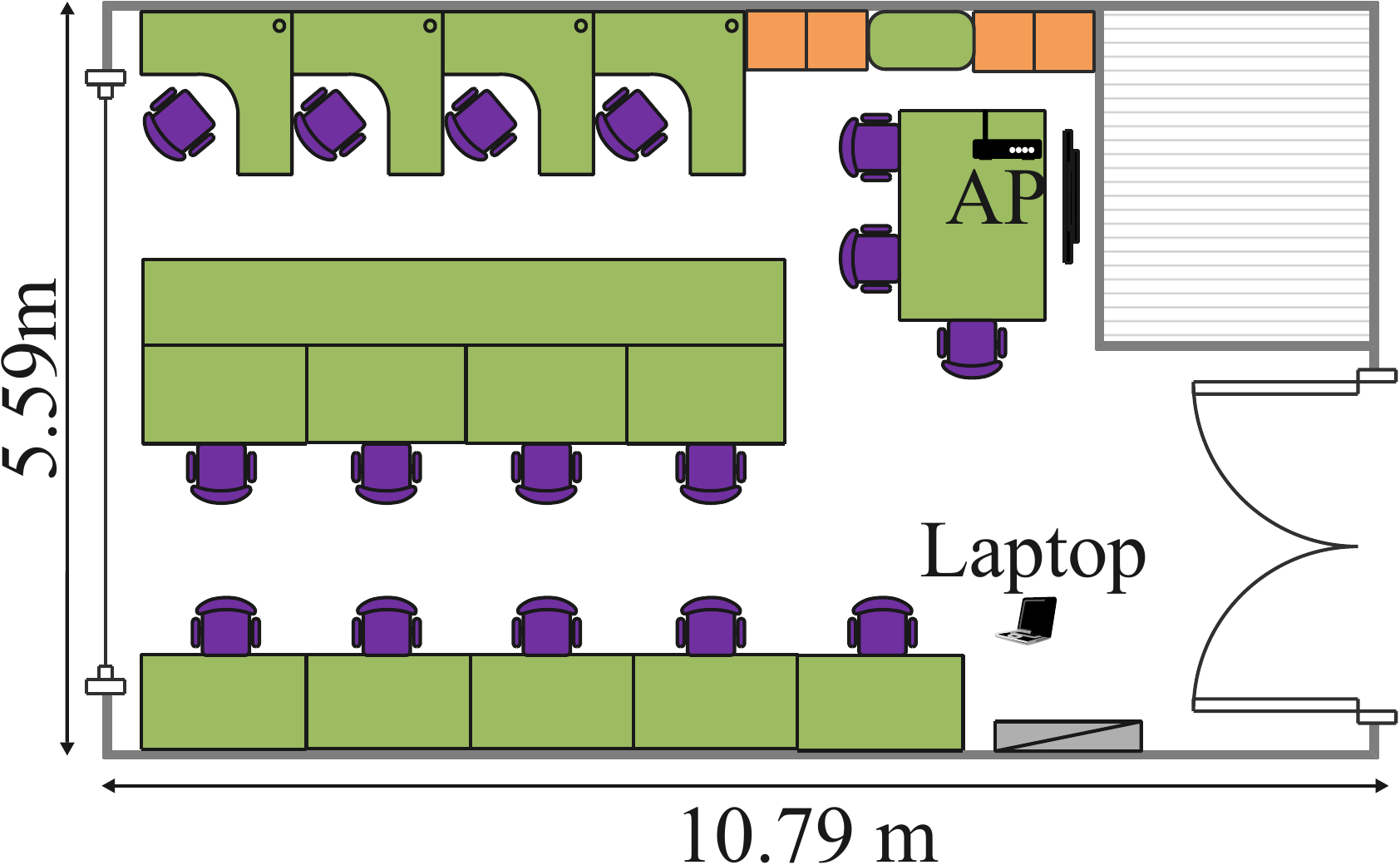}
                    
			\end{minipage}
			\label{subfig:office}
		}
        \vspace{-1ex}
        \subfloat[The classroom.]{
            \hspace{-2ex}
		  \begin{minipage}[b]{0.49\linewidth}
		        \centering
			    \includegraphics[width = \textwidth, height = 0.59\textwidth]{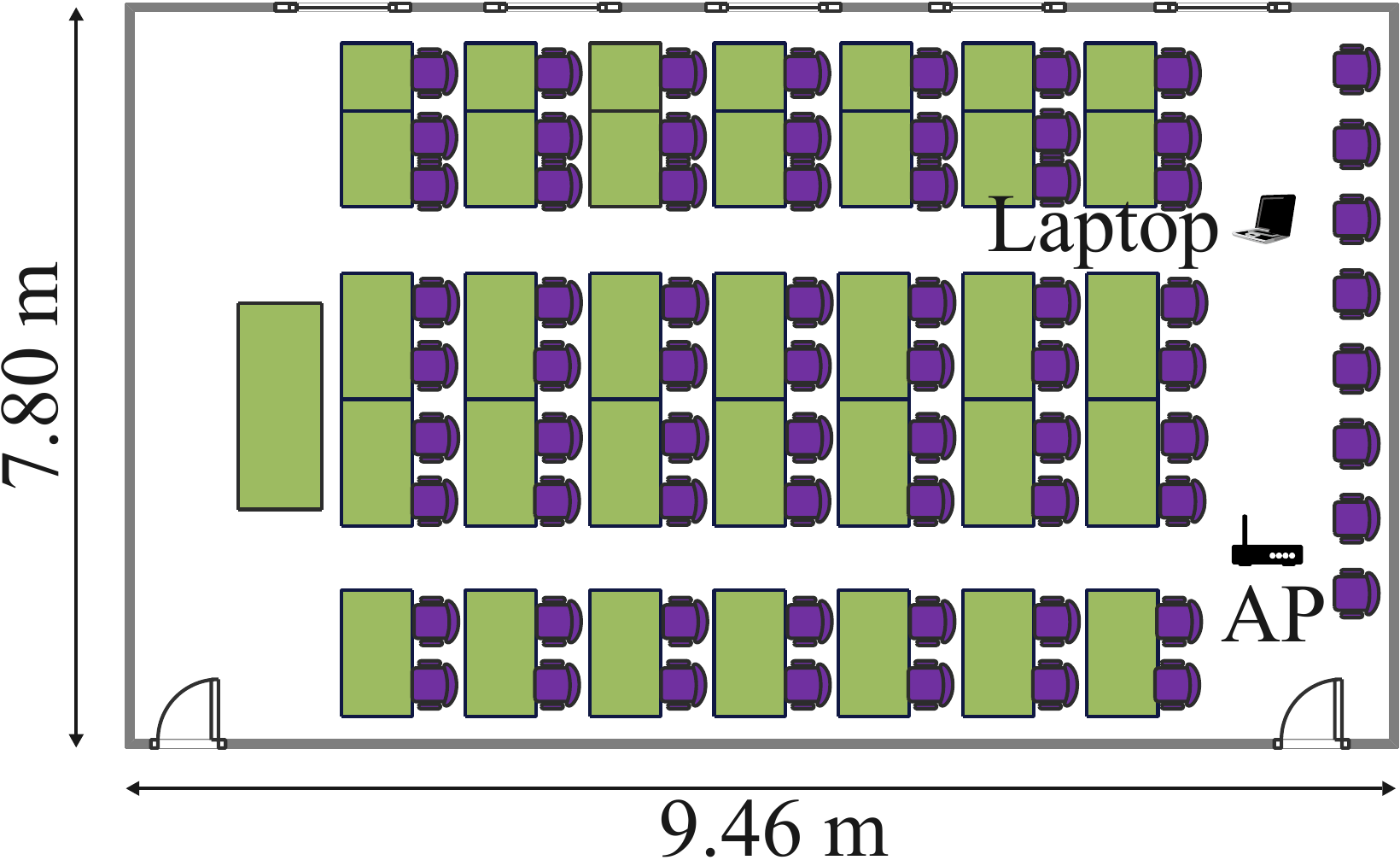}
                    
			\end{minipage}
			\label{subfig:classroom}
		}
        \subfloat[The corridor.]{
		    \begin{minipage}[b]{0.49\linewidth}
		        \centering
			    \includegraphics[width = \textwidth, height = 0.59\textwidth]{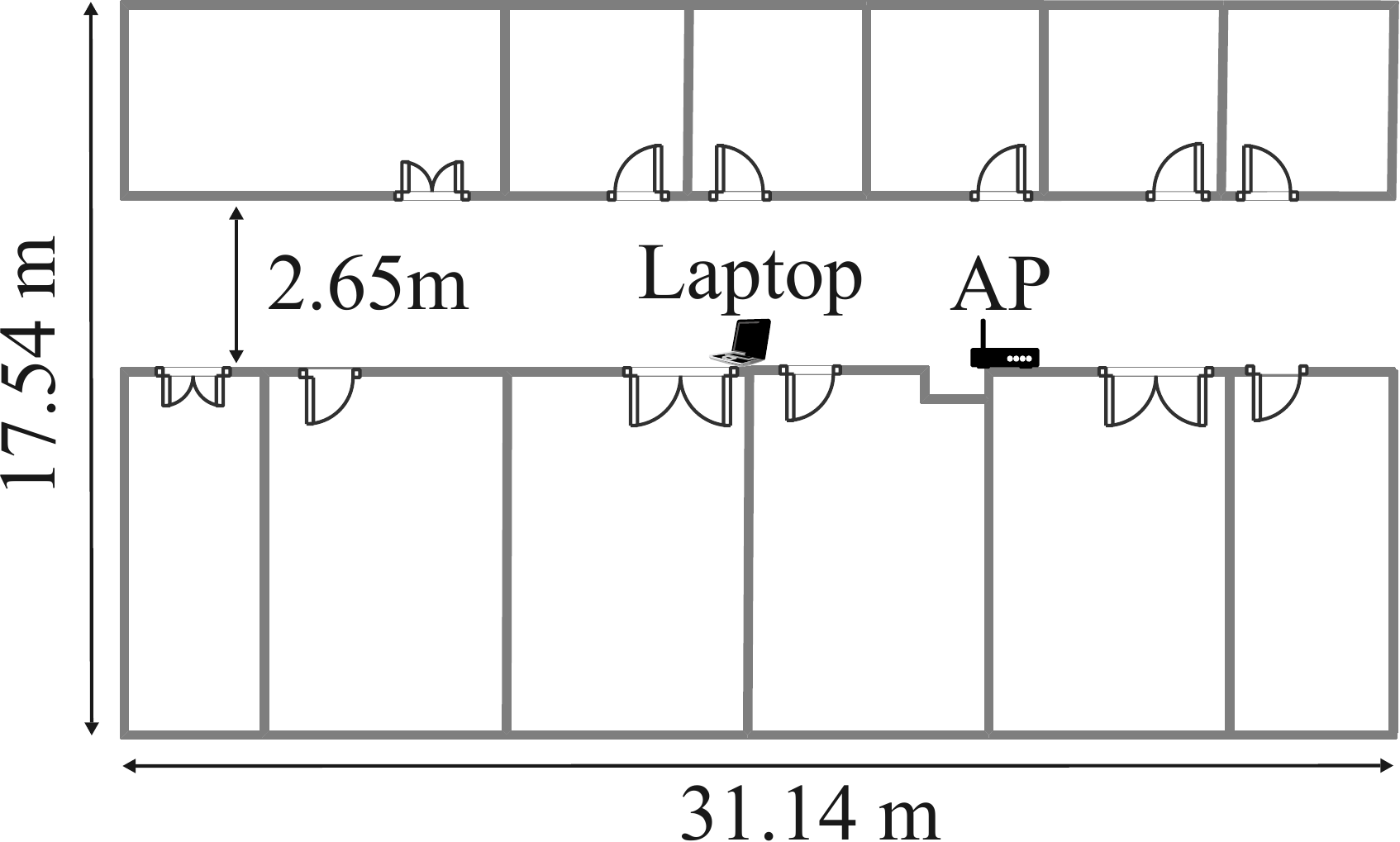}
                    
			\end{minipage}
			\label{subfig:corridor}
		}            
		\caption{The layout of four selected experimental scenes.}
		\label{fig:scenes}
\end{figure}
We deploy the proposed \name in four typical real-world indoor environments to comprehensively evaluate its performance. Fig.~\ref{fig:scenes} shows the layout of the four scenarios: a meeting room (3.53 $\times$ 2.85~\!m$^2$), an laboratory (10.79 $\times$ 5.59~\!m$^2$), a classroom (9.46 $\times$ 7.8~\!m$^2$), and a corridor (31.14 $\times$ 2.65~\!m$^2$). The ASUS RT-AX86U router is used as the AP for CSI acquisition. The AP runs at a center frequency of $f$=5~\!GHz with a bandwidth of 20~\!MHz. We only use 20~\!MHz out of the 160~\!MHz total bandwidth to show \name's accuracy for legacy wireless infrastructures. The CSI across 128 OFDM subcarriers are captured by the AX-CSI tool~\cite{ax-csi} from three detachable antennas with 8.9~\!cm spacing. In Fig.~\ref{fig:scenes}, we also illustrate the placements of Tx and Rx across different scenes. We employ a single AP with three antennas and one common laptop in the device-free setup. The line-of-sight (LoS) distance between them is 4~\!m (3.5~\!m in the meeting room) and their height is fixed at 1.2~\!m. In the device-based settings, we only use a single AP with three antennas for localization. This is a readily available configuration in typical indoor environments, demonstrating the practicality of \name. 
The height and LoS distance of transceivers are fixed across all experiments unless otherwise specified. We report the localization error defined as the distance between the estimated position provided by localization models and the actual ground truth label obtained from the real world.

\subsection{Baseline Selection}
To evaluate the performance of \name, we compare it with six state-of-the-art CSI-based localization systems, three selected for device-free settings and three for device-based settings. 
In particular, the selected device-free approaches are introduced as follows:
\begin{itemize}
    \item Widar2.0~\cite{widar2} is a device-free wireless localization system that enables passive human localization with a single pair of transceivers.
    \item WiTraj~\cite{WiTraj} employs the ratio of CSI from two antennae of each RX to enhance the quality of estimated signal characteristics, thereby achieving more accurate localization.
    \item DSCP~\cite{dscp} is a deep convolutional neural network (CNN)-based device-free localization system that divides the indoor scene into different areas to collect the amplitudes differences as the fingerprint database.  
\end{itemize}
We also select the following device-based baselines:
\begin{itemize}
    \item SpotFi~\cite{spotfi} computes the AoA with super-resolution algorithms and identifies the AoA of the direct path between the localization target and AP to deliver an accurate localization performance.
    \item $M^3$~\cite{m3} exploits a multipath-assisted approach to enable single AP localization through the combination of azimuths and the relative ToF of LoS signal and reflection signals.
    \item CiFi~\cite{CiFi} extracts phase data of CSI as input to a deep CNN to train the CNN model in the offline phase. The object's location is predicted based on the trained CNN and newly produced phase data.
\end{itemize}

\subsection{Performance of Scene Model Construction}
\begin{figure}[t]
    \setlength\abovecaptionskip{6pt}
    \vspace{-2ex}
	\captionsetup[subfigure]{justification=centering}
		\centering
		\subfloat[Learning curves for relative permittivities.]{
        \hspace{-2ex}
		  \begin{minipage}[b]{0.49\linewidth}
		        \centering
			    \includegraphics[width = \textwidth]{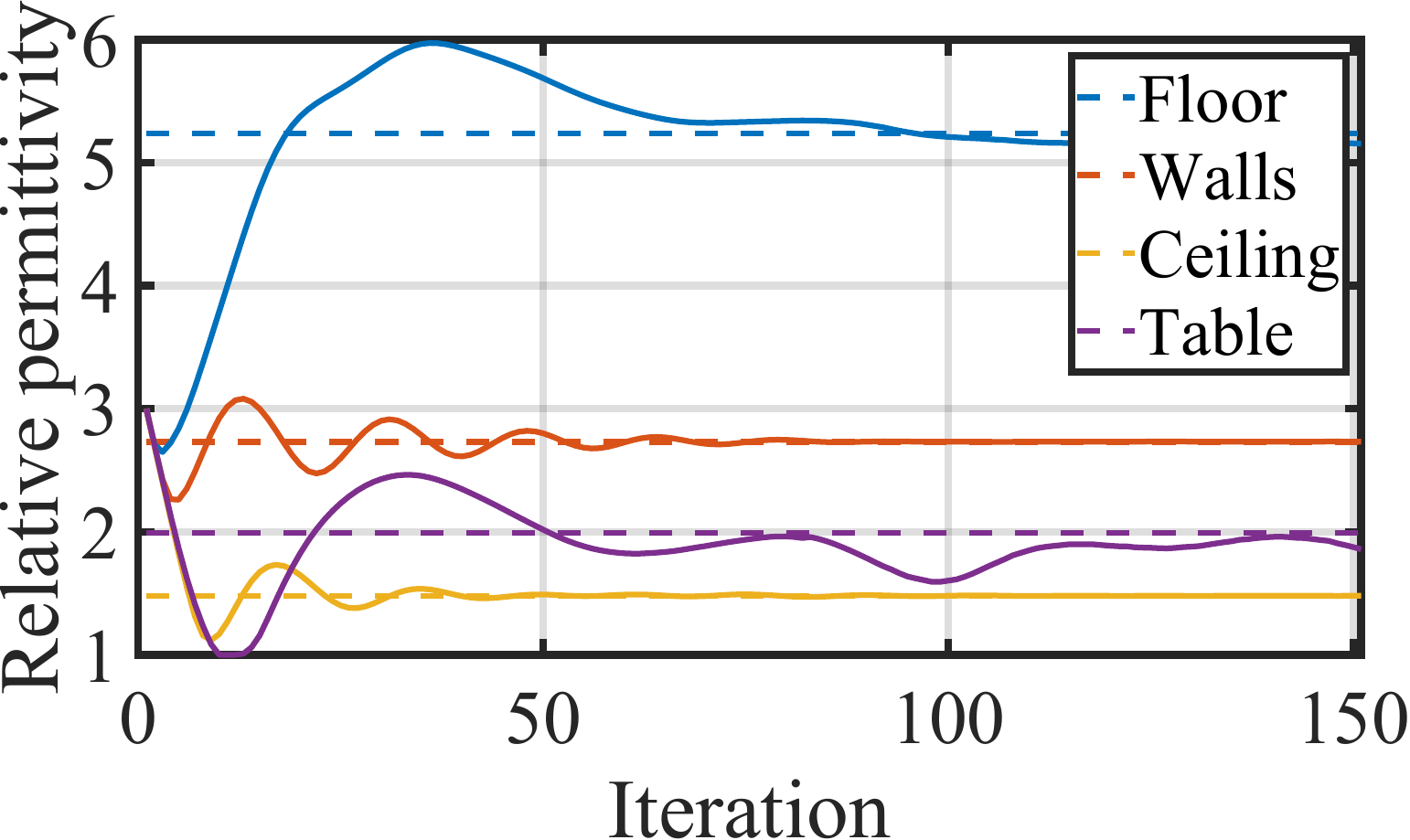}                   
			\end{minipage}
			\label{subfig:calibration-em-1}
		}
		\subfloat[Learning curves for conductivities.]{
		    \begin{minipage}[b]{0.49\linewidth}
		        \centering
			    \includegraphics[width = \textwidth]{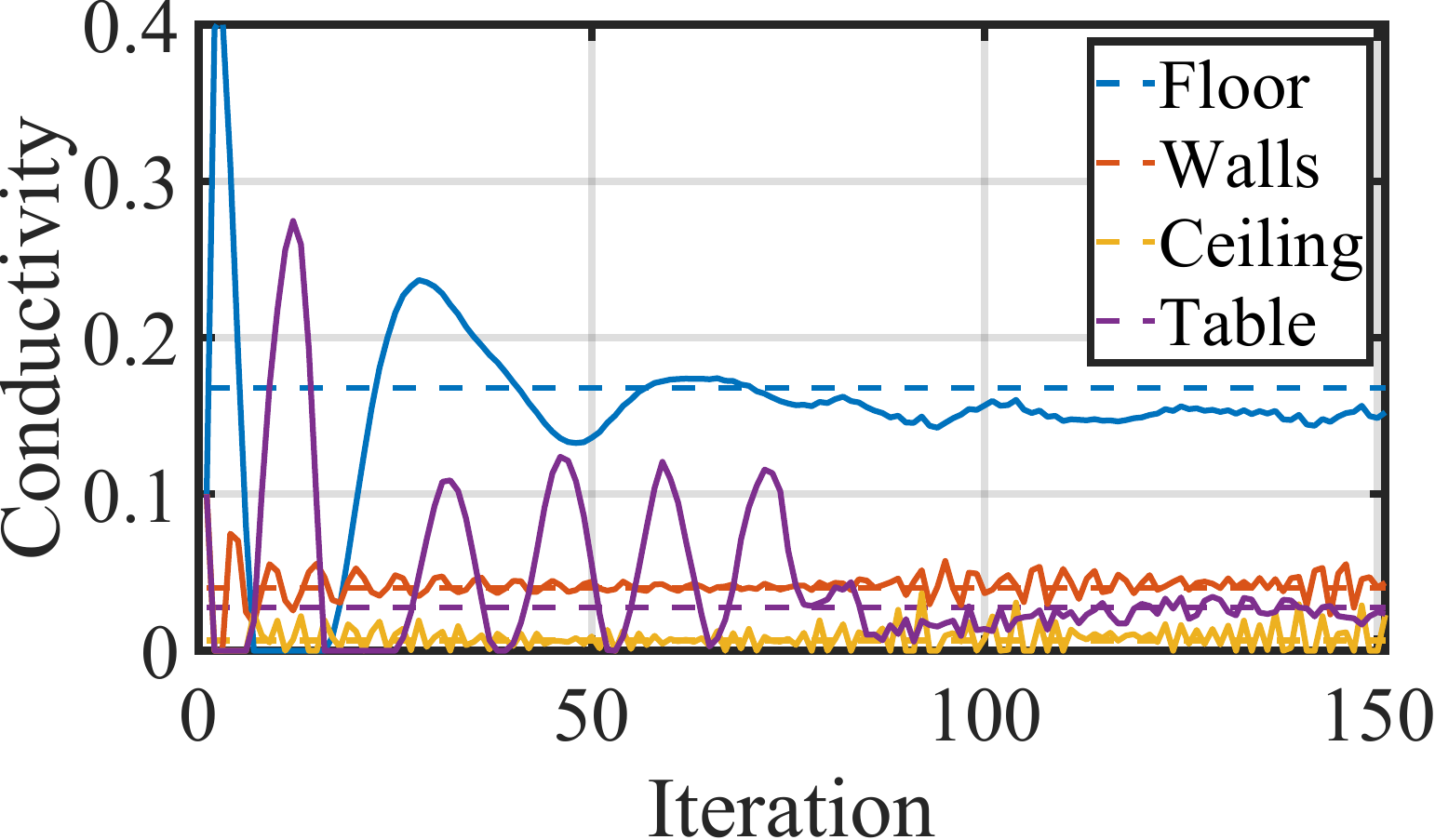}
                    
			\end{minipage}
			\label{subfig:calibration-em-2}
		}
        \vspace{-1ex}
        \subfloat[Learning curves for positions.]{
            \hspace{-2ex}
		  \begin{minipage}[b]{0.49\linewidth}
		        \centering
			    \includegraphics[width = \textwidth]{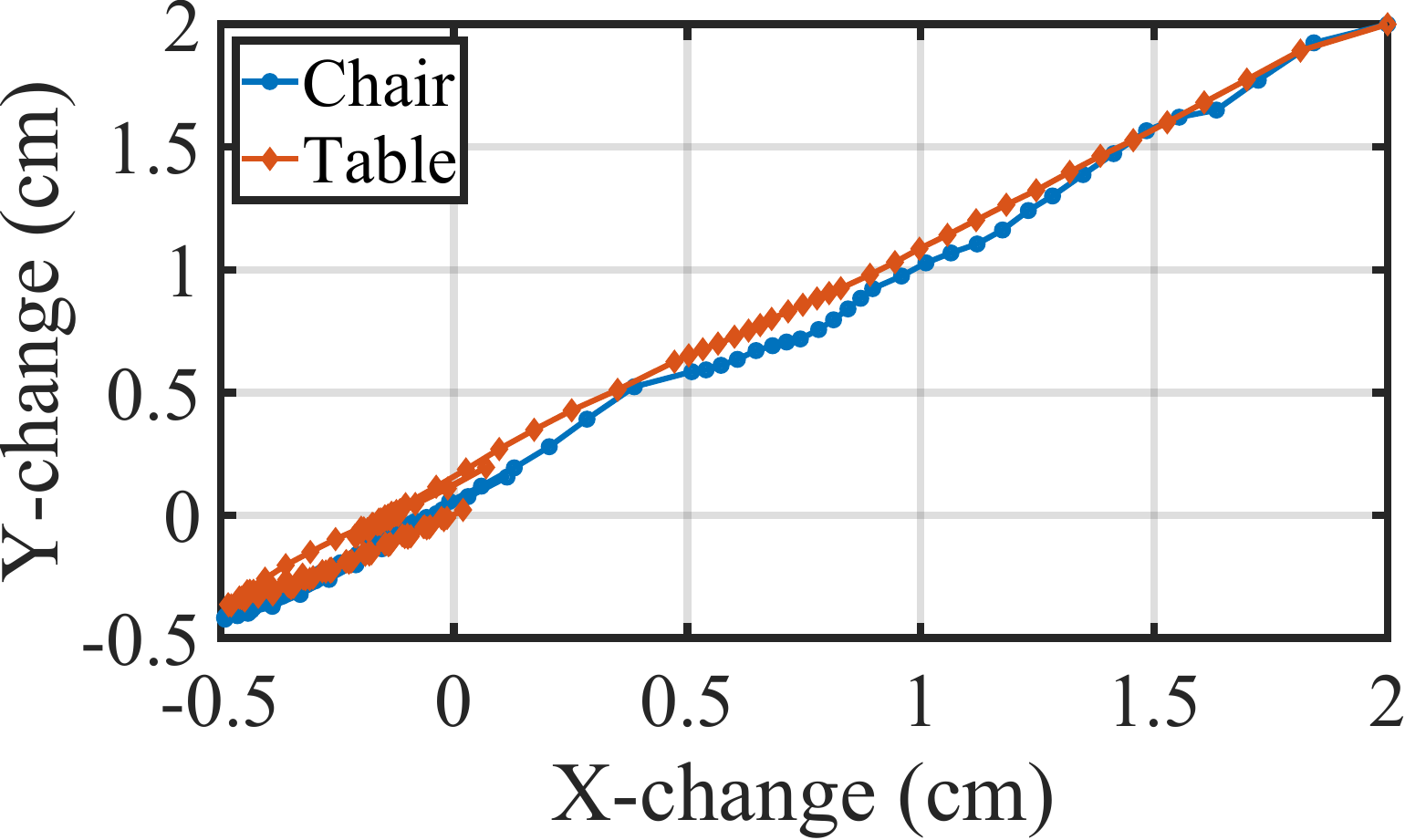}
                    
			\end{minipage}
			\label{subfig:calibration-pos}
		}
        \subfloat[CSI differences.]{
		    \begin{minipage}[b]{0.49\linewidth}
		        \centering
			    \includegraphics[width = \textwidth]{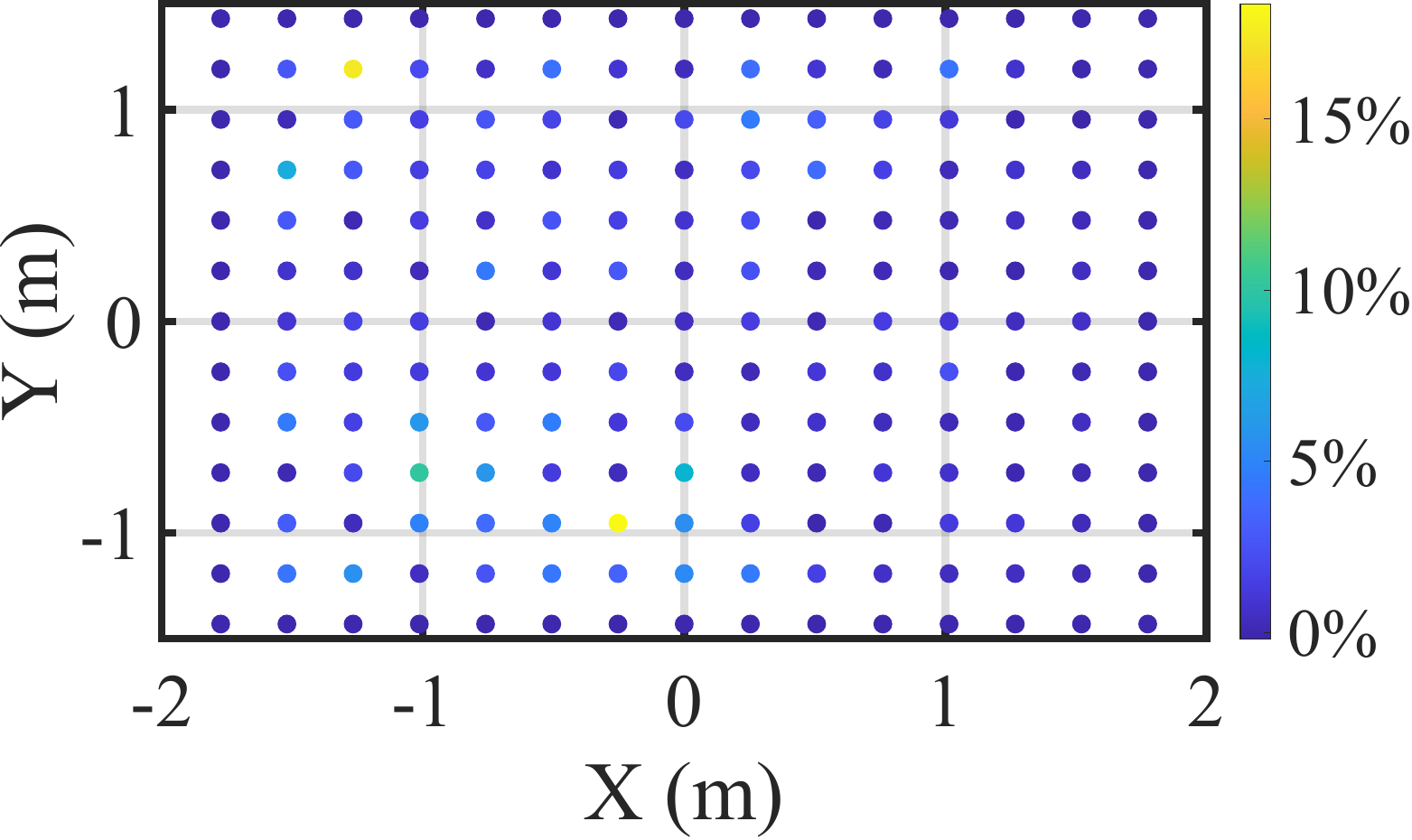}
                    
			\end{minipage}
			\label{subfig:csi}
		}            
		\caption{Learning curves for the scene parameters and the CSI before and after calibration.}
		\label{fig:calibration}	
\end{figure}
Before diving into evaluations of the localization performance, we conduct extensive experiments to show the effectiveness of the proposed scene model construction approach.
We initially establish a ground-truth scene model for the meeting room depicted in Fig.~\ref{subfig:conference}, and designate 195 measurement locations within it to collect CSI. Subsequently, we randomly alter the scene parameters of the DT, including the EM properties of the wall, ceiling, and floor and the position of the meeting table and chairs. With the measured CSI, we observe whether background model construction could accurately reproduce these values. The construction processes are illustrated in Fig.~\ref{subfig:calibration-em-1}, \ref{subfig:calibration-em-2}, and \ref{subfig:calibration-pos}. We can clearly see that these scene parameters start to converge to the ground truth after 50 iterations. All their values match the correct values very closely after 150 iterations. 
To verify the effectiveness of high-fidelity background model construction, we also conduct the experiment in real-world scenes. 
Taking the meeting room as an example, we collect CSI data entries from predefined 143 measurement points to form a dataset. We utilize the dataset to calibrate the scene parameters of the meeting room. We show the difference between the real-world CSI and the corresponding CSI obtained in the calibrated digital copy of the meeting room in Fig.~\ref{subfig:csi}. We can find that it is possible to recover these scene parameters via gradient descent with the measured CSI from the meeting room.

\subsection{Accuracy of Localization}
In this section, we evaluate the localization accuracy achieved by \name in different configurations to test its superiority and robustness.

\subsubsection{Overall Performance}\label{over-performance}
\begin{figure}[b]
    \setlength\abovecaptionskip{6pt}
    \vspace{-2ex}
	\captionsetup[subfigure]{justification=centering}
		\centering
		\subfloat[Device-free \& LoS setting.]{
        \hspace{-2ex}
		  \begin{minipage}[b]{0.49\linewidth}
		        \centering
			    \includegraphics[width = \textwidth, height = 0.59\textwidth]{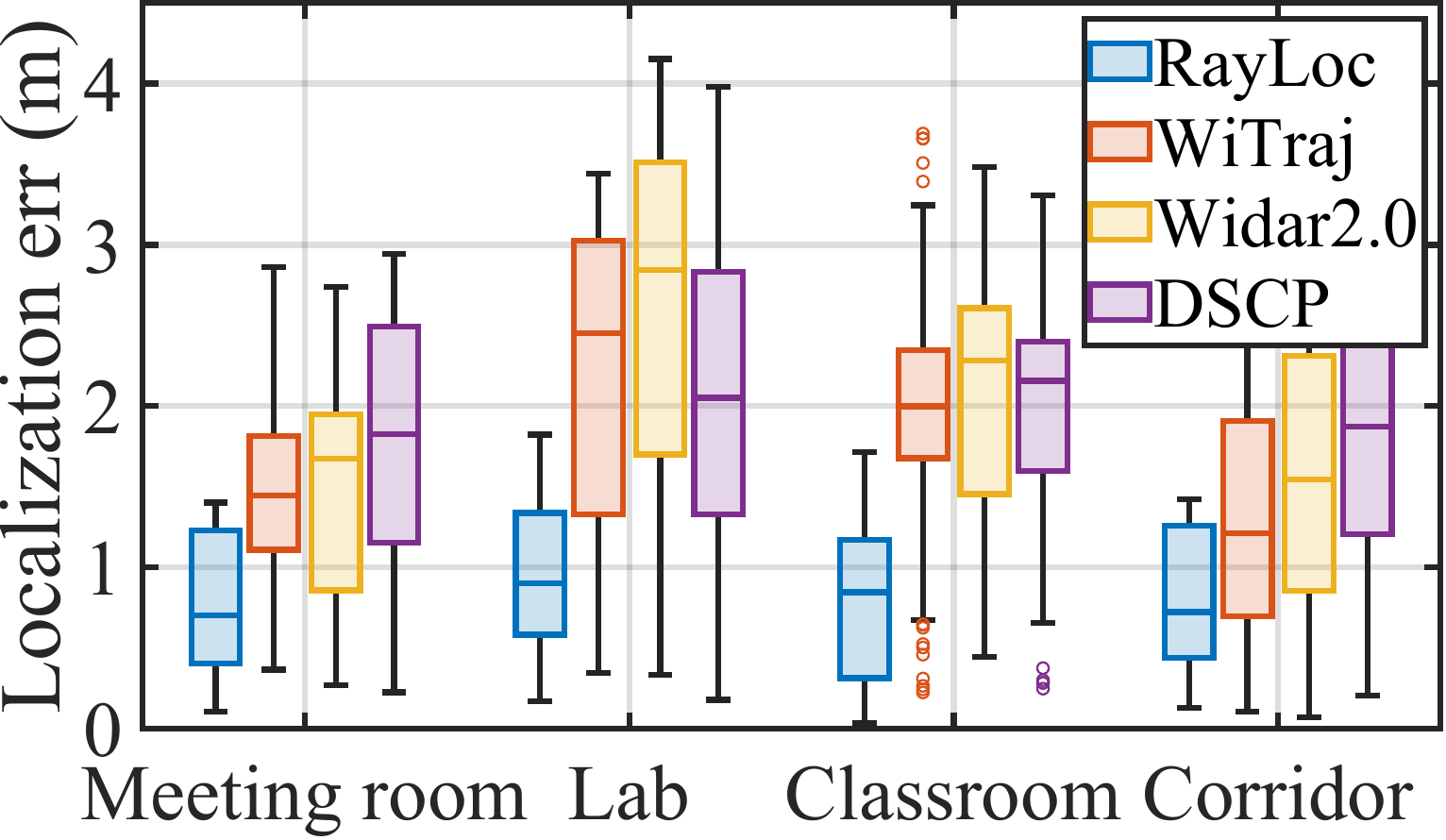}                   
			\end{minipage}
			\label{subfig:free-los}
		}
		\subfloat[Device-based \& LoS setting.]{
		    \begin{minipage}[b]{0.49\linewidth}
		        \centering
			    \includegraphics[width = \textwidth, height = 0.59\textwidth]{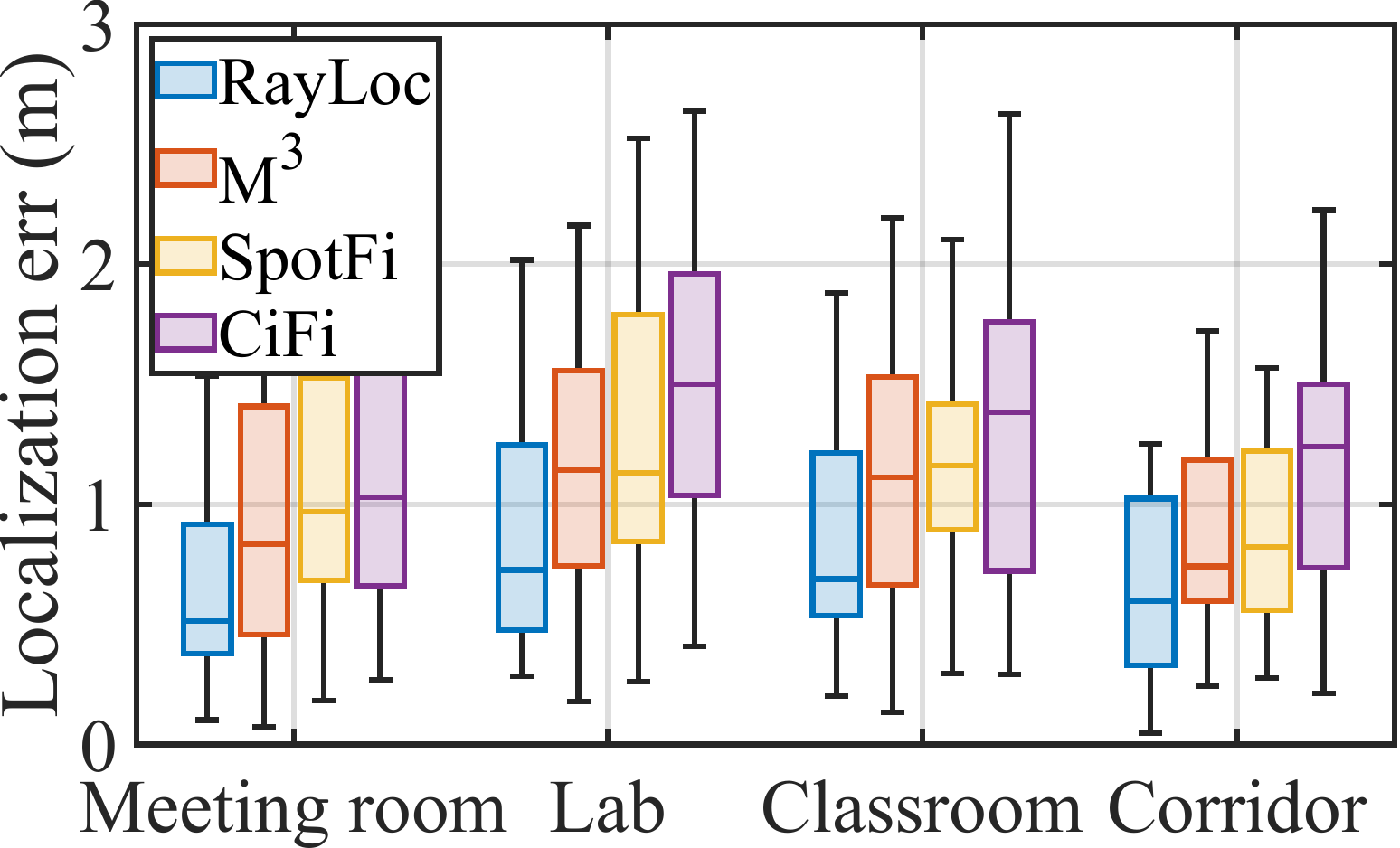}
                    
			\end{minipage}
			\label{subfig:based-los}
		}
        \vspace{-1ex}
        \subfloat[Device-free \& NLoS setting.]{
            \hspace{-2ex}
		  \begin{minipage}[b]{0.49\linewidth}
		        \centering
			    \includegraphics[width = \textwidth, height = 0.59\textwidth]{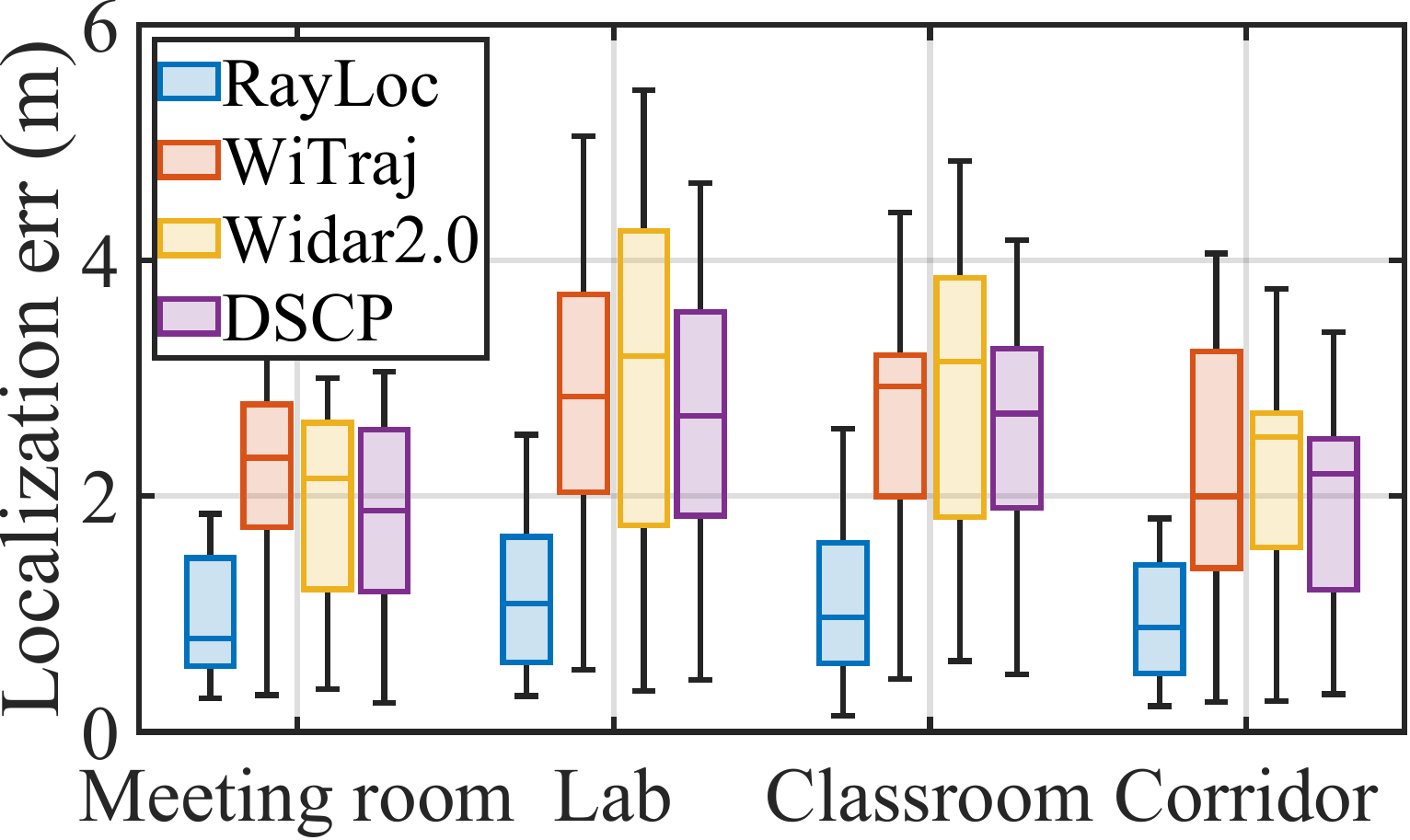}
                    
			\end{minipage}
			\label{subfig:free-nlos}
		}
        \subfloat[Device-based \& NLoS setting.]{
		    \begin{minipage}[b]{0.49\linewidth}
		        \centering
			    \includegraphics[width = \textwidth, height = 0.59\textwidth]{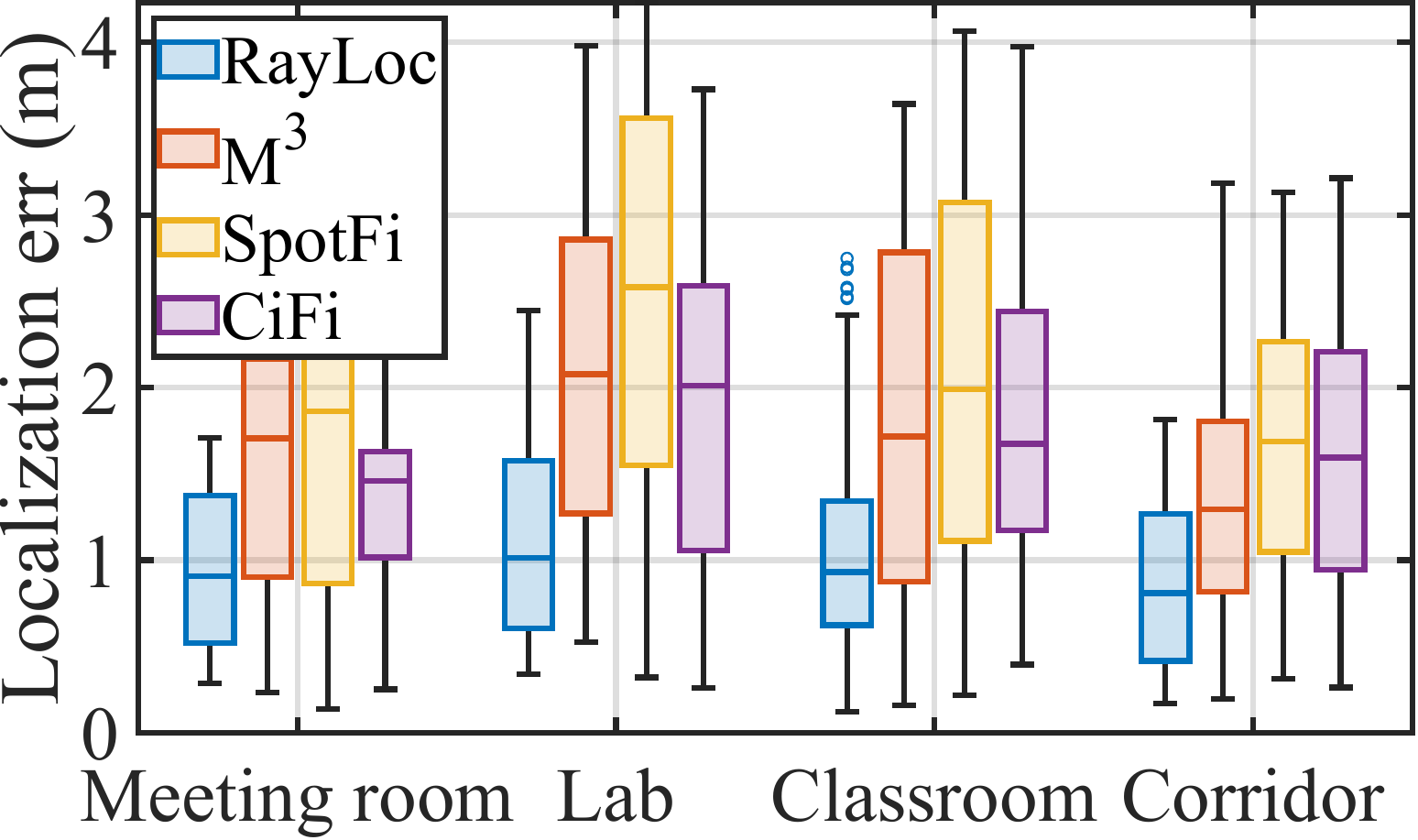}
                    
			\end{minipage}
			\label{subfig:based-nlos}
		}            
		\caption{Localization errors for device-free and device-based settings in LoS and NLoS scenarios.}
		\label{fig:overall}	
\end{figure}
To evaluate the performance of \name, we compare it with various state-of-the-art methods across different experimental scenes. 
In each scene, we measure the localization error of \name and all the selected baselines in the LoS and NLoS scenarios. The NLoS means the LoS path cannot be obtained due to the blockage of obstacles in indoor environments, generally considered a more challenging scenario for localization tasks. To create the NLoS scenario, we lower the height of the AP to 0.2~\!m. Since most indoor obstacles are higher than 0.2~\!m, they will block the LoS path. Specifically, we place some obstacles, including chairs and wooden boards, in the corridor.
We repeat the experiments across 50 different locations in all the experimental scenes. 
The results under device-free and device-based settings are detailed in Fig.~\ref{fig:overall}. 
Our evaluation shows that \name can outperform the considered baselines across all the experimental scenes in the LoS and NLoS scenarios, respectively.

Fig.~\ref{subfig:free-los} shows the localization errors achieved by \name and device-free baselines in the selected four indoor scenes. Fig.~\ref{subfig:free-los} reveals that compared to other approaches, \name reduces the average localization errors by over 0.72~\!m, 1.09~\!m, 1.17~\!m, and 0.55~\!m, in the meeting room, laboratory, classroom, and corridor, respectively. The performance improvement of \name can be attributed to its innovative ``turn foes into friends'' approach, which leverages environment knowledge (e.g., multipath) as valuable localization information. In contrast, conventional localization methods treat these same environment factors as interference that degrades their accuracy. Fig.~\ref{subfig:based-los} shows the localization errors of these device-based localization methods. From Fig.~\ref{subfig:based-los}, we can observe that device-based approaches outperform device-free counterparts and \name achieves nearly 0.44~\!m improvement in localization accuracy across all four scenes. The reason behind the observation is that a calibrated scene model can generate more stable and accurate measurement values of wireless CSI, benefiting precise localization. 

Fig.~\ref{subfig:free-nlos} and Fig.~\ref{subfig:based-nlos} show the localization errors achieved by these localization methods in device-free and device-based settings, respectively. Due to the absence of LoS, we can see that the errors for all localization methods in NLoS settings are higher compared to LoS scenarios. However, we can also observe that even in the challenging NLoS configuration, \name still outperforms the other localization methods. The effectiveness can be attributed to the fact that, after creating and calibrating a digital replica of the scene, \name has more prior knowledge of the environment. Combined with the gradient-descent optimization, it allows the \name to accurately locate the object. 

\subsubsection{Performance under Different Clutter Levels}
\begin{figure}[h]
	\captionsetup[subfigure]{justification=centering}
		\centering
		\subfloat[Device-free settings.]{
        \hspace{-2ex}
		  \begin{minipage}[b]{0.49\linewidth}
		        \centering
			    \includegraphics[width = \textwidth]{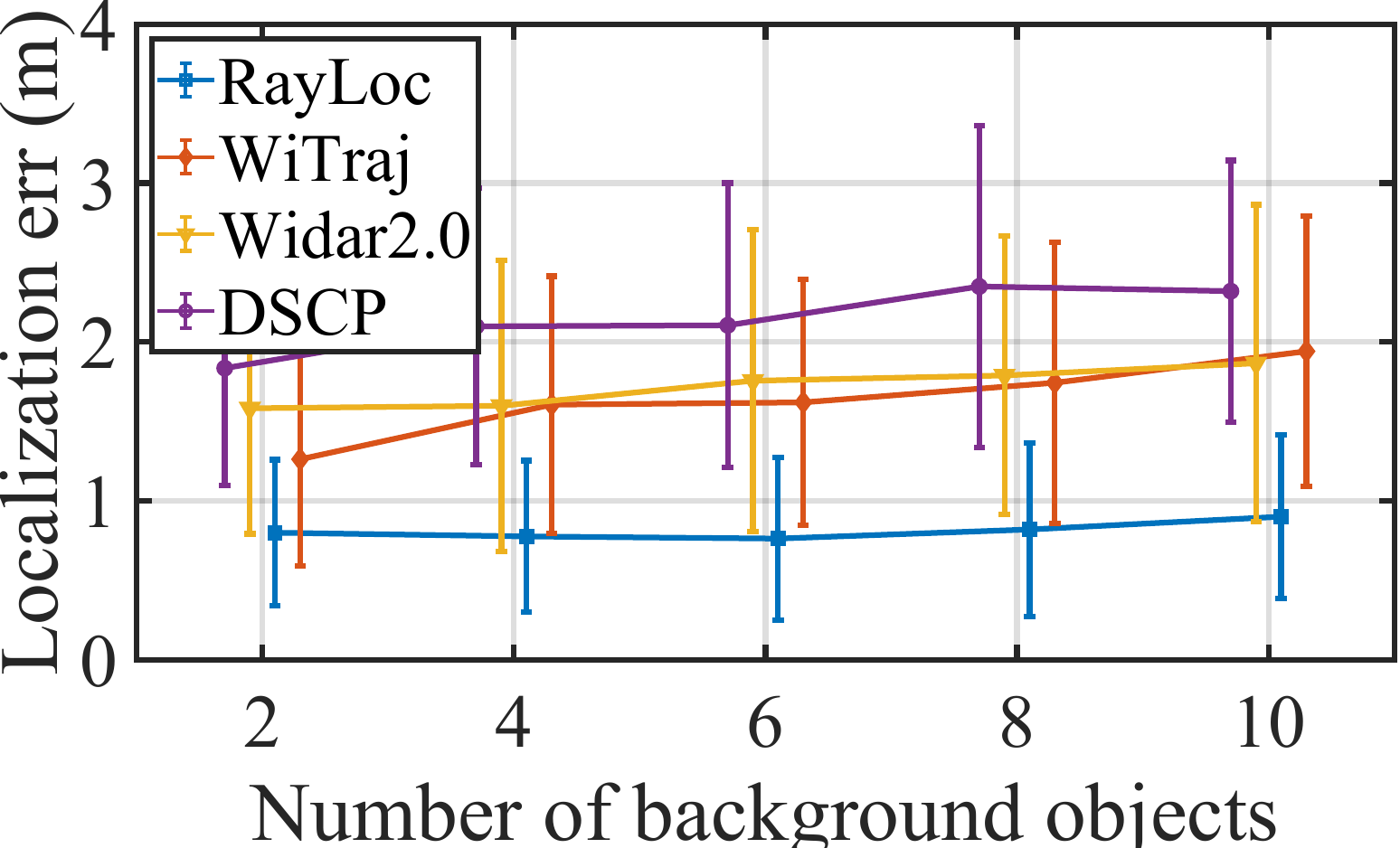}                   
			\end{minipage}
			\label{subfig:clutter-free}
		}
		\subfloat[Device-based settings.]{
		    \begin{minipage}[b]{0.49\linewidth}
		        \centering
			    \includegraphics[width = \textwidth]{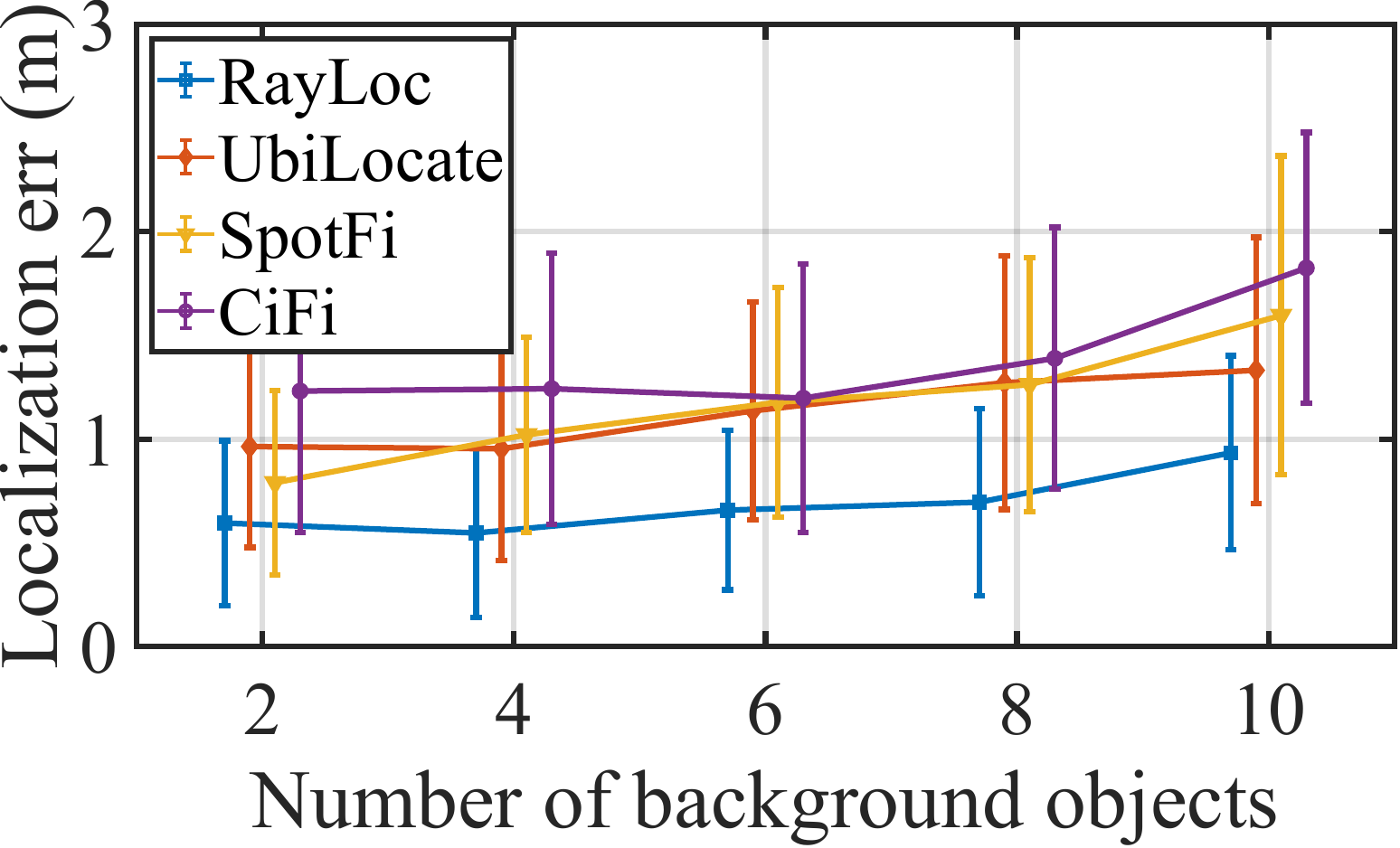}
                    
			\end{minipage}
			\label{subfig:clutter-based}
		}
		\caption{Performance under different background clutter levels.}
		\label{fig:clutter}	
\end{figure}

We now evaluate the overall localization accuracy of various approaches under different background clutter level. In the experiment, clutter level refers to the number of objects and the degree of messiness in an indoor space. To this end, we place different cardboard boxes (height: 50~\!cm; width: 65~\!cm; thickness: 13~\!cm) in the corridor, specifically arranging two and ten boxes to test the localization errors of these systems. The results are illustrated in Fig.~\ref{fig:clutter}. It is readily apparent that the superiority of \name becomes more evident as the clutter level increases, whether under device-free or device-based settings. This is because current localization methods treat objects in the scene as harmful, whereas \name incorporates prior knowledge of the environment into the system, effectively converting complex environment into beneficial information for localization.

\subsection{Impact of Practical Factors}
In this section, we further explore the performance of \name with various potential factors, including the distance between transceivers, AP height, AP number, and related hyper-parameters.
\subsubsection{Impact of Distance}
\begin{figure}[h]
    \vspace{-3ex}
	\captionsetup[subfigure]{justification=centering}
		\centering
		\subfloat[Device-free settings.]{
        \hspace{-2ex}
		  \begin{minipage}[b]{0.49\linewidth}
		        \centering
			    \includegraphics[width = \textwidth]{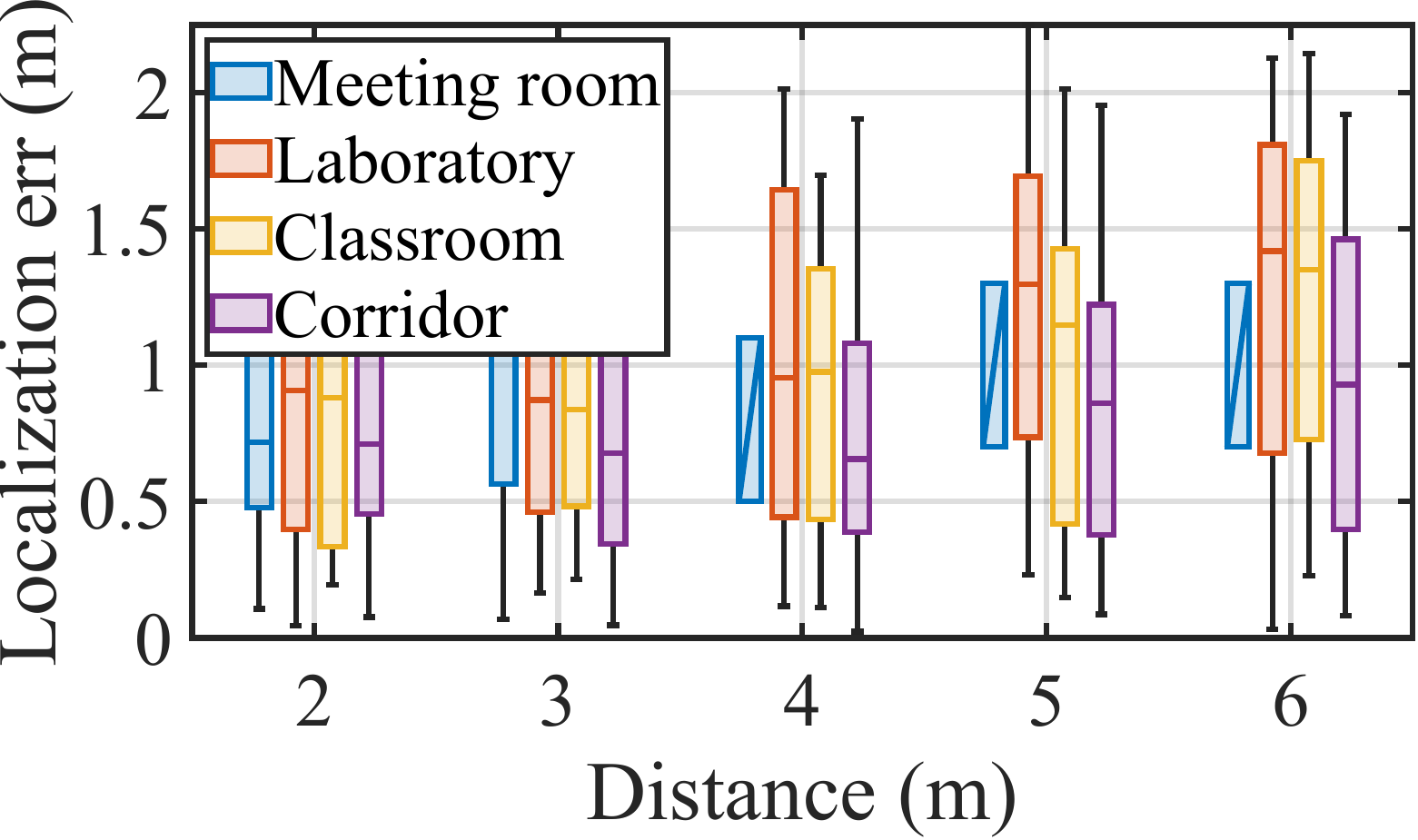}                   
			\end{minipage}
			\label{subfig:distance-free}
		}
		\subfloat[Device-based settings.]{
		    \begin{minipage}[b]{0.49\linewidth}
		        \centering
			    \includegraphics[width = \textwidth, height = 0.59\textwidth]{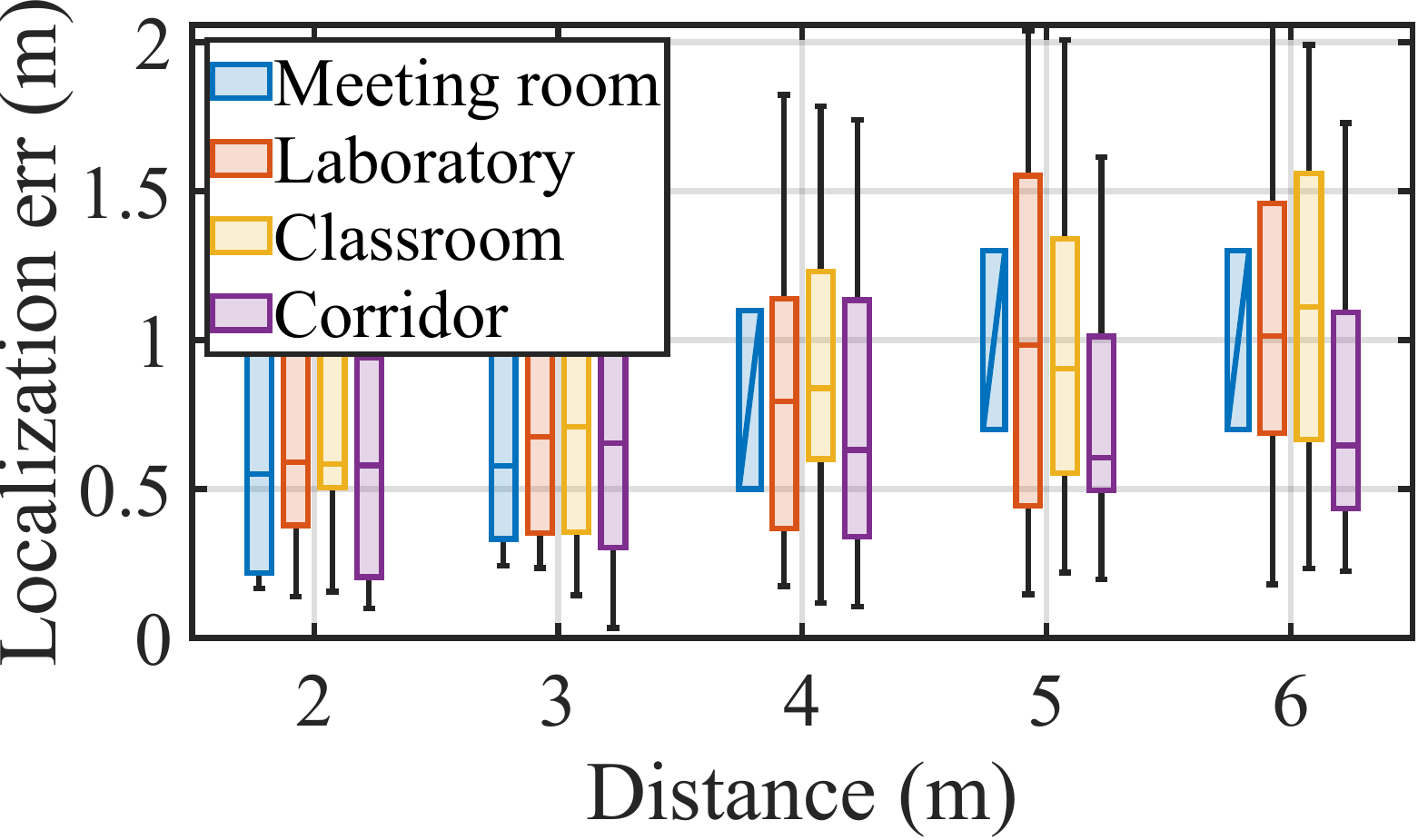}
                    
			\end{minipage}
			\label{subfig:distance-based}
		}
		\caption{Impact of the distance between the transceivers.}
        \vspace{-3ex}
		\label{fig:distance}	
\end{figure}

To evaluate the robustness of \name, we first investigate the impact of the distance between transceivers on the localization error. The distance between transceivers in the set of experiments refers to the separation between the AP and the laptop in a device-free configuration and the distance between the AP and the target device in a device-based setup. We move the AP to vary the distance ranging from 2~\!m to 6~\!m across the four scenes.
The localization errors under the device-free and device-based settings are shown in Fig.~\ref{subfig:distance-free} and Fig.~\ref{subfig:distance-based}, respectively. It is clear to see that the localization error increases with the growing distance, in both device-free and device-based configurations. The diminishing effectiveness of \name can be explained by the fact that as the distance increases, CSI becomes more susceptible to environmental noise, resulting in instability and a consequent rise in localization error. Nevertheless, the fact that the localization error remains below 0.4~\!m as the distance increases to 6~\!m verifies the robustness of \name.
\subsubsection{Impact of AP Height}
\begin{figure}[h]
    \vspace{-3ex}
	\captionsetup[subfigure]{justification=centering}
		\centering
		\subfloat[Device-free settings.]{
        \hspace{-2ex}
		  \begin{minipage}[b]{0.49\linewidth}
		        \centering
			    \includegraphics[width = \textwidth]{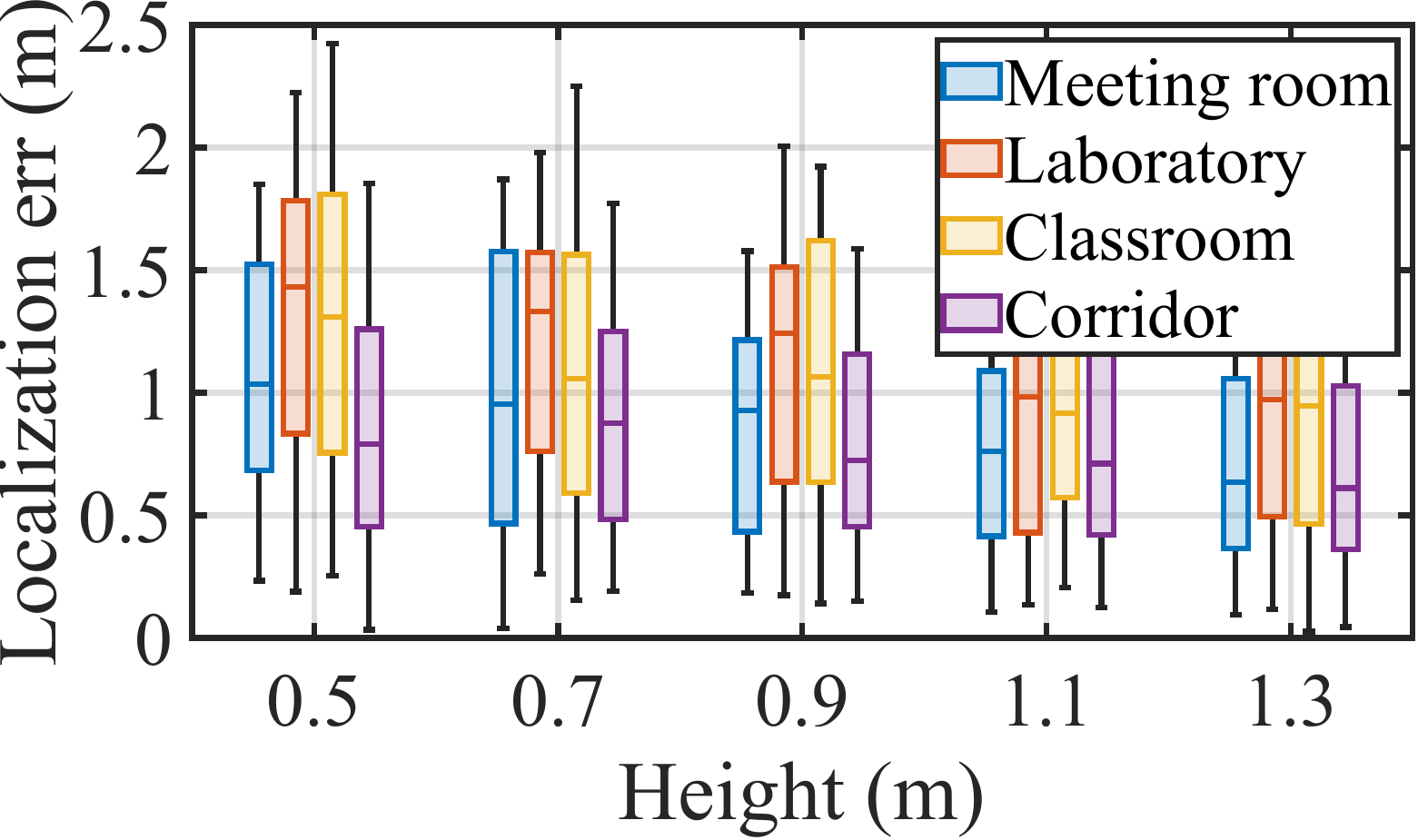}                   
			\end{minipage}
			\label{subfig:height-free}
		}
		\subfloat[Device-based settings.]{
		    \begin{minipage}[b]{0.49\linewidth}
		        \centering
			    \includegraphics[width = \textwidth]{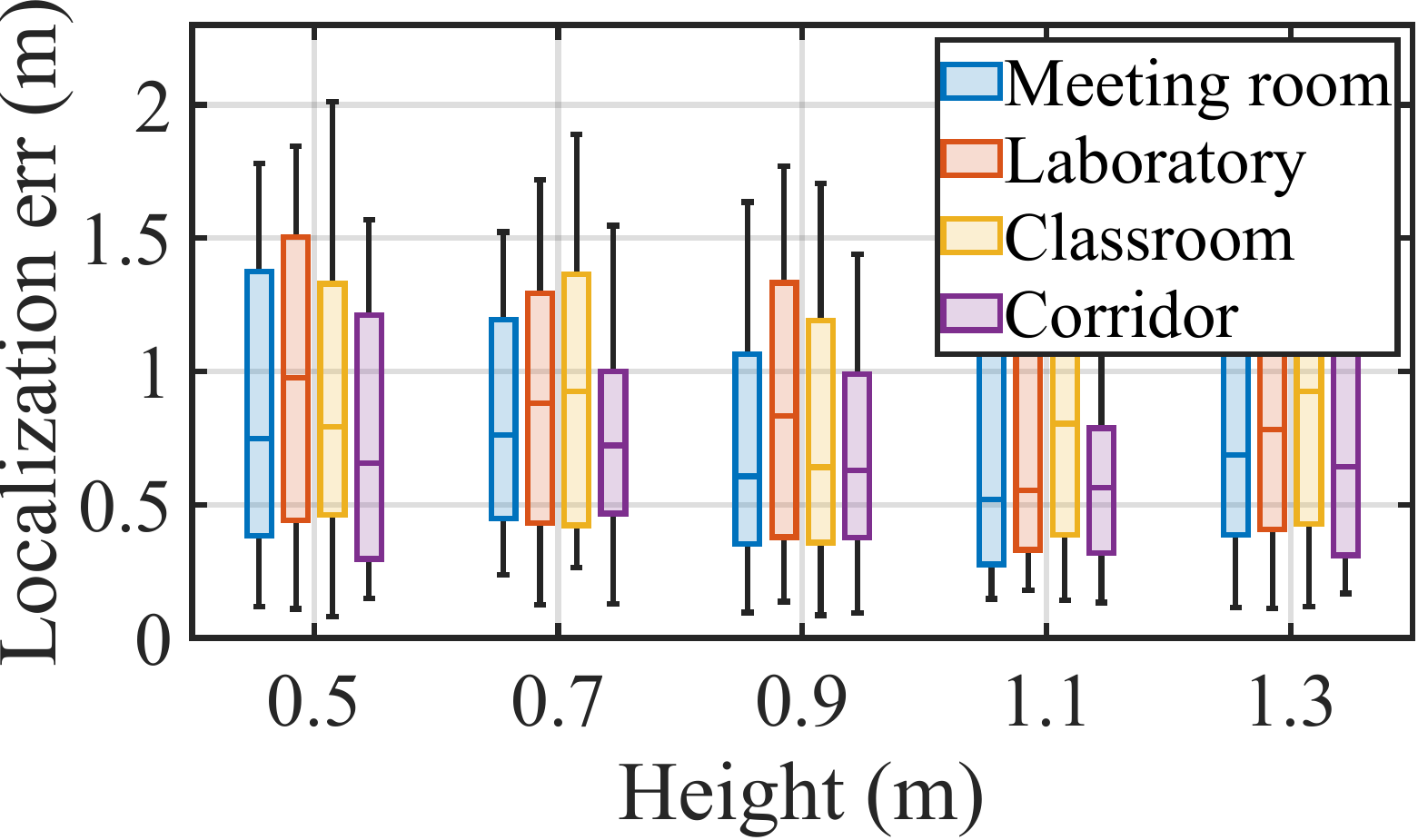}
                    
			\end{minipage}
			\label{subfig:height-based}
		}
		\caption{Impact of AP height.}
        \vspace{-3ex}
		\label{fig:height}	
\end{figure}

In practical situations, the height of AP can vary significantly. For instance, APs are typically placed on tables or cabinets, which can result in different heights. The height at which the AP is positioned greatly affects the precision of localization. Therefore, we evaluate the localization errors of \name by varying the height of the AP in the set of experiments. We vary the height of the AP from 0.5~\!m to 1.3~\!m with a step size of 0.2~\!m. Fig.~\ref{subfig:height-free} and Fig.~\ref{subfig:height-based} illustrate the localization errors for each height in device-free and device-based settings. In device-free configurations, lower AP heights result in decreased localization accuracy due to increased obstacles and potential NLoS scenarios. In device-based settings, when the target device and the AP are at a similar height (1~\!m), the localization error is the smallest. As the absolute height difference between AP and the target device increases, so does the localization error. However, changes in AP height only lead to a maximum increase of 0.5~\!m in \name’s localization error, demonstrating its robustness to height variations.

\subsubsection{Impact of AP number}
\begin{figure}[h]
    \vspace{-3ex}
	\captionsetup[subfigure]{justification=centering}
		\centering
		\subfloat[Device-free settings.]{
        \hspace{-2ex}
		  \begin{minipage}[b]{0.49\linewidth}
		        \centering
			    \includegraphics[width = \textwidth, height = 0.59\textwidth]{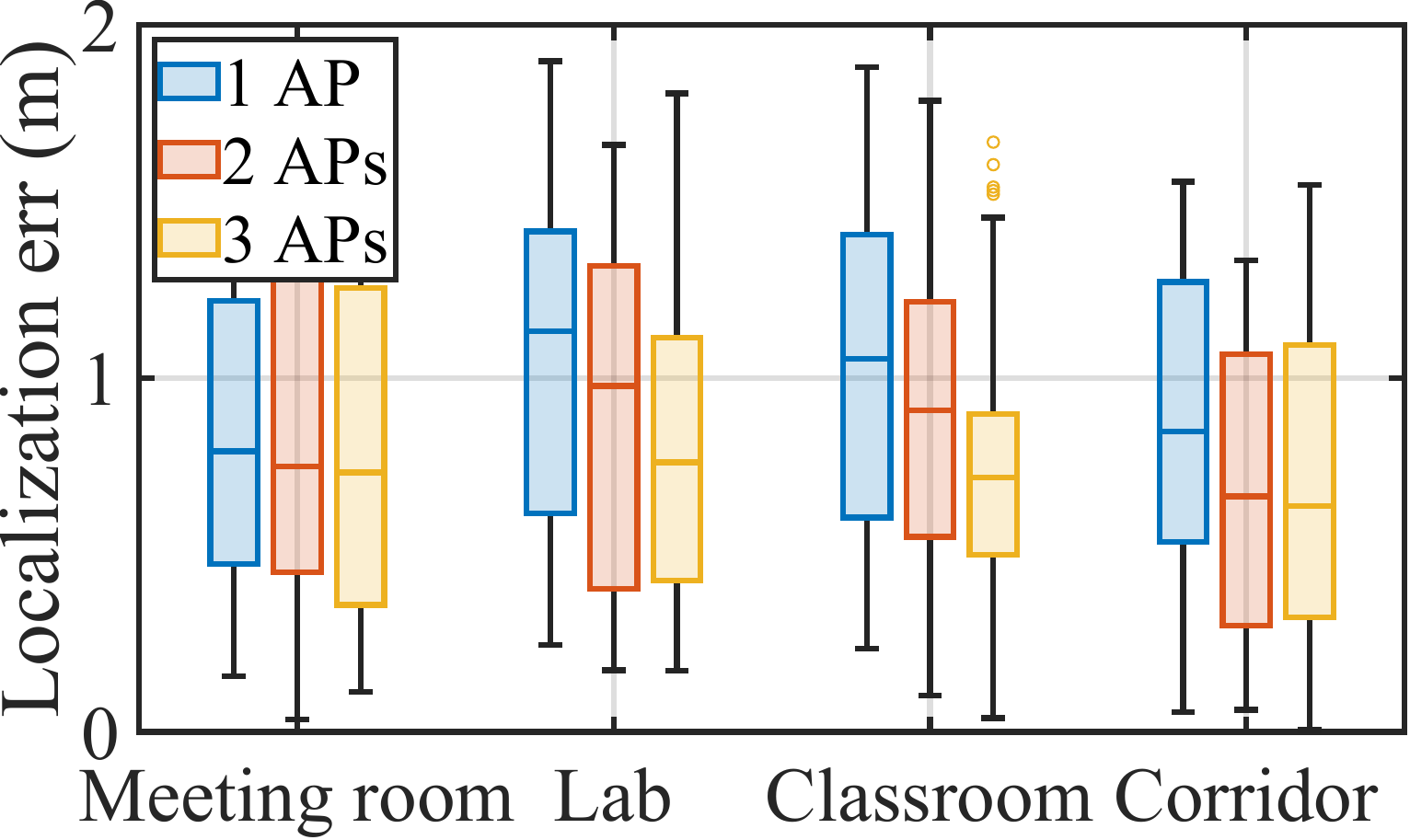}                   
			\end{minipage}
			\label{subfig:ap-free}
		}
		\subfloat[Device-based settings.]{
		    \begin{minipage}[b]{0.49\linewidth}
		        \centering
			    \includegraphics[width = \textwidth, height = 0.59\textwidth]{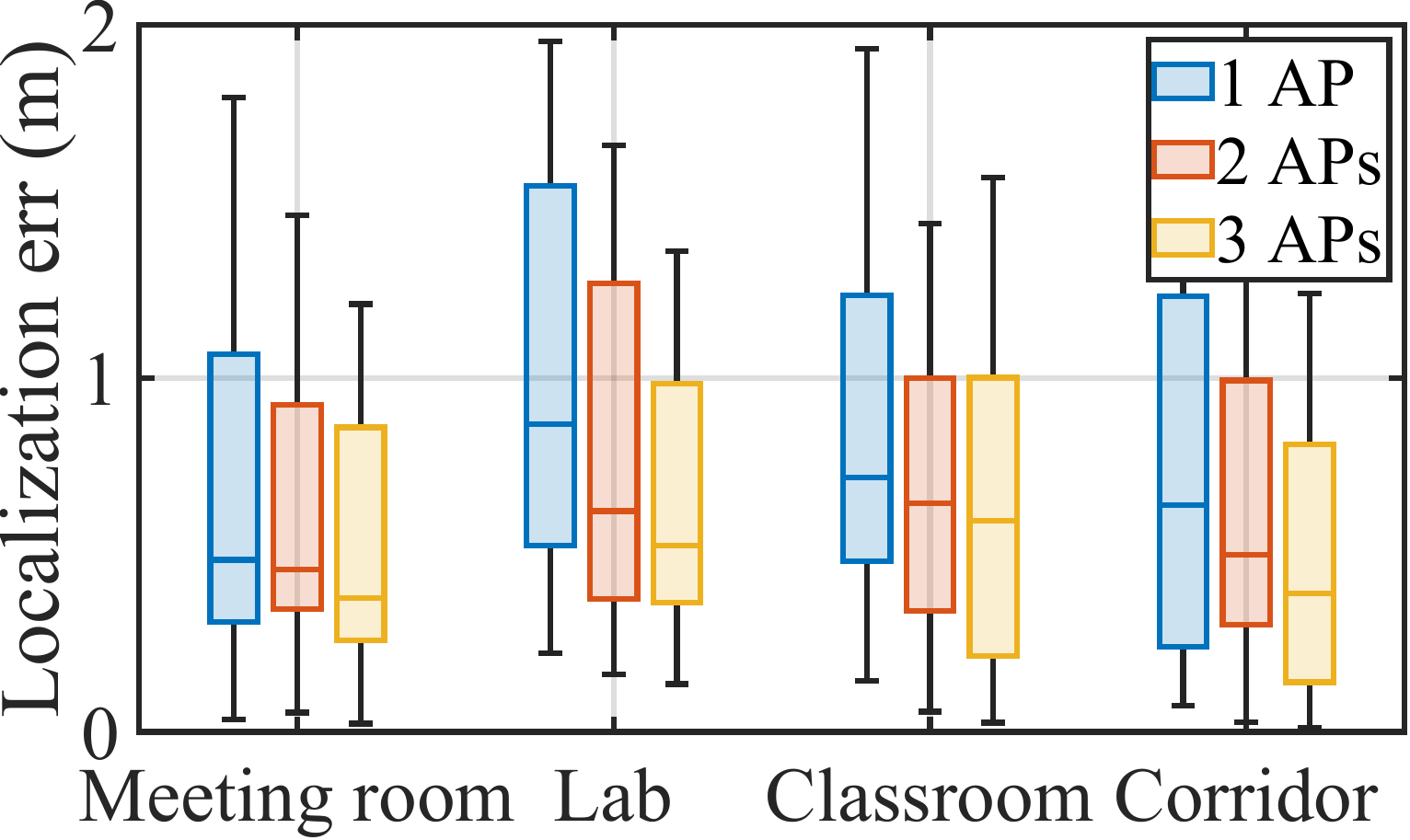}
                    
			\end{minipage}
			\label{subfig:ap-based}
		}
		\caption{Impact of AP number.}
        \vspace{-3ex}
		\label{fig:ap}	
\end{figure}

More APs can obtain more CSI measurements, which also means a more comprehensive characterization of the environment. Hence, in the subsequent set of experiments, we evaluate the impact of varying numbers of APs on localization error to investigate the robustness of \name to AP densities. We vary the AP number from one to three with a step size of one. The localization errors of \name for device-free and device-based configurations are reported in Fig.~\ref{subfig:ap-free} and Fig.~\ref{subfig:ap-based}, respectively. As shown in Fig.~\ref{fig:ap}, we can see that the localization error decreases with an increasing number of APs in all four scenes, as expected. The improvement in localization accuracy demonstrates the scalability and robustness of \name.

\subsubsection{Impact of Variance $\sigma$}
\begin{figure}[h]
    \vspace{-2ex}
	\captionsetup[subfigure]{justification=centering}
		\centering
		\subfloat[Device-free settings.]{
        \hspace{-3ex}
		  \begin{minipage}[b]{0.49\linewidth}
		        \centering
			    \includegraphics[width = \textwidth]{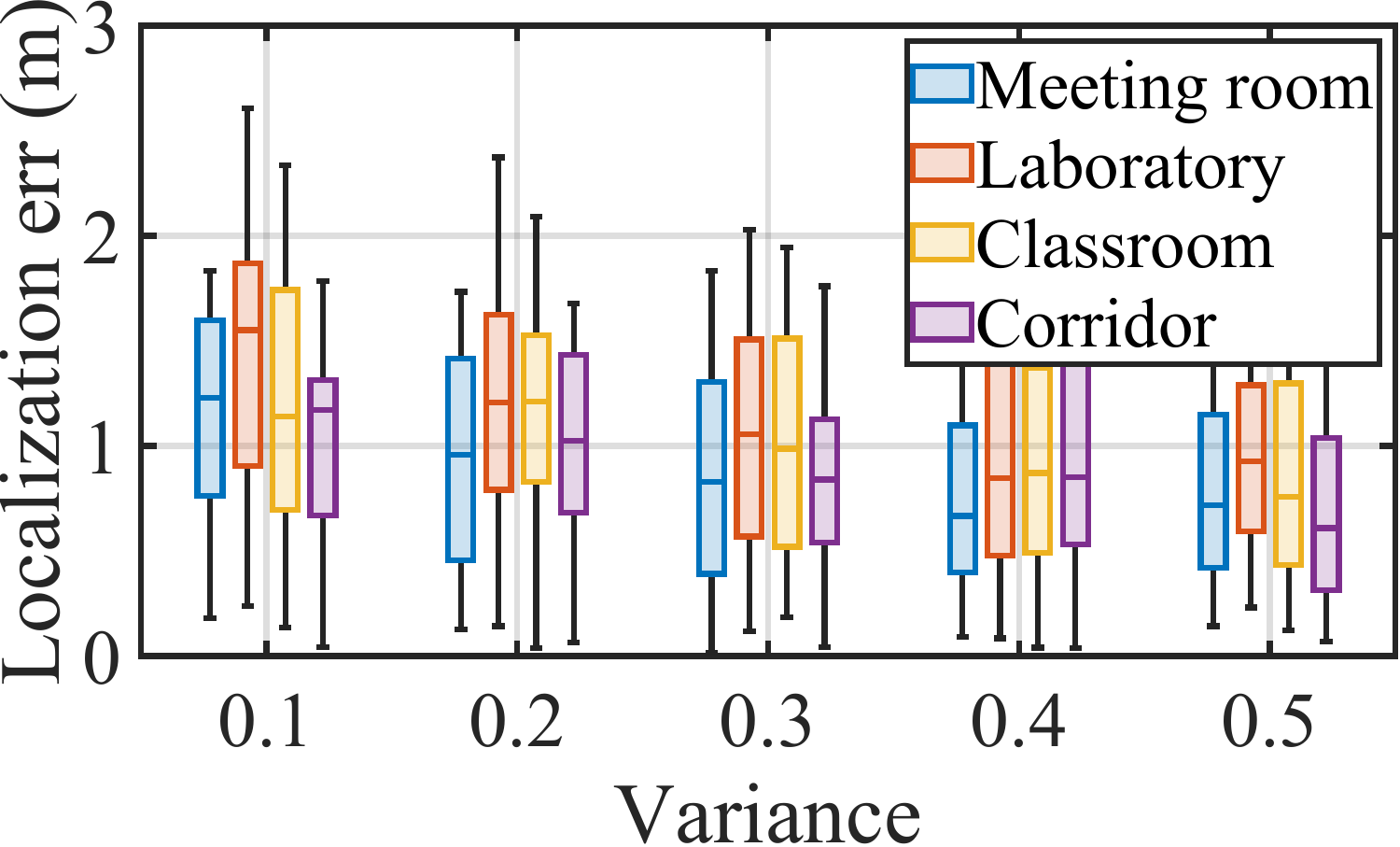}                   
			\end{minipage}
			\label{subfig:var-free}
		}
		\subfloat[Device-based settings.]{
		    \begin{minipage}[b]{0.49\linewidth}
		        \centering
			    \includegraphics[width = \textwidth, height = 0.59\textwidth]{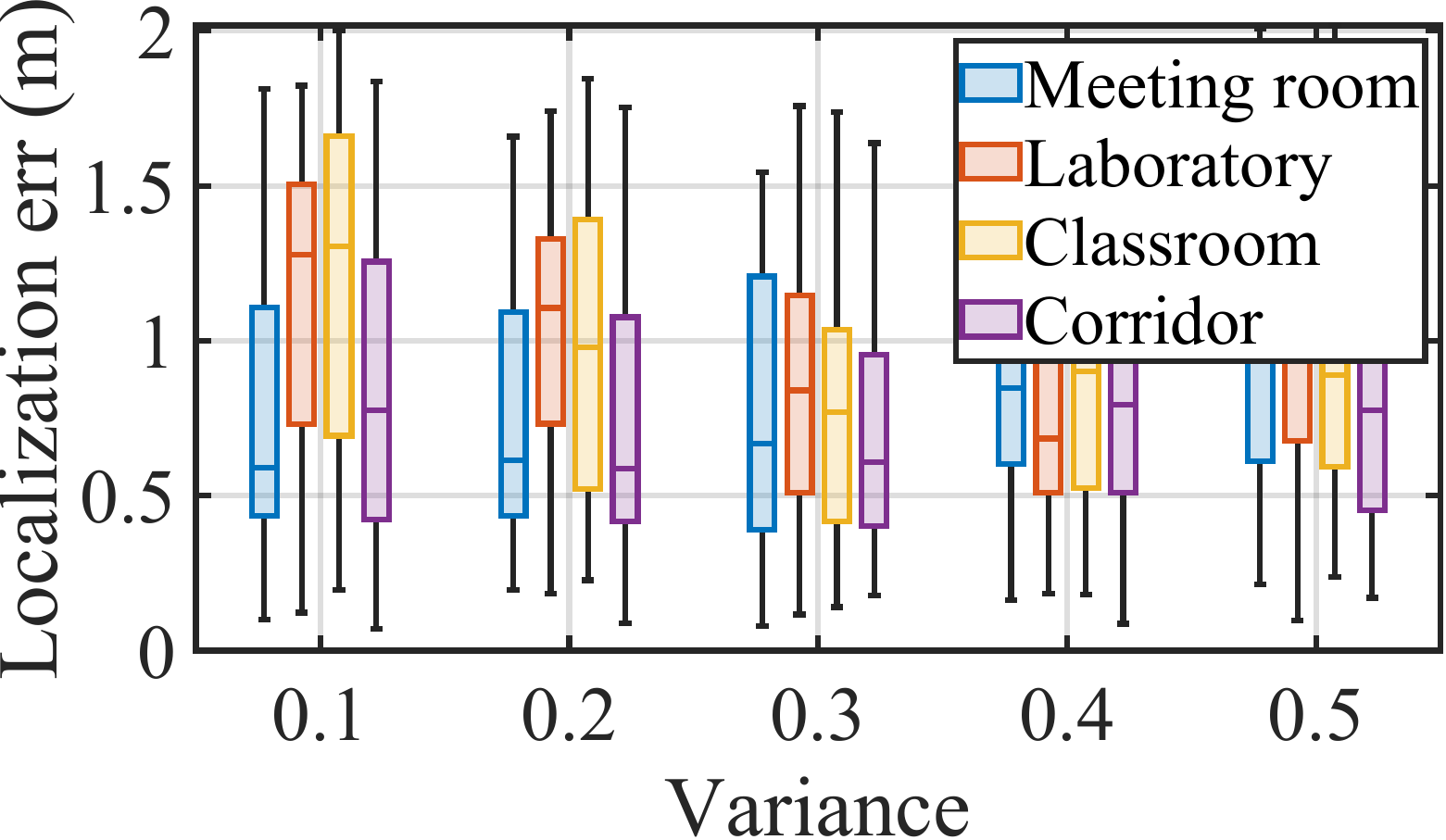}
                    
			\end{minipage}
			\label{subfig:var-based}
		}
		\caption{Impact of variance $\sigma$.}
        \vspace{-3ex}
		\label{fig:var-pos}	
\end{figure}

The role of the variance $\sigma$ in \name is to adjust the relative importance of the sampling positions, determining the proportion that each sample contributes to the final gradient calculation. A larger value $\sigma$ causes the gradient to emphasize global information, while a smaller $\sigma$ focuses more on the local area around the current position. Excessive $\sigma$ can lead to unstable optimization and prevent convergence, whereas a smaller value of $\sigma$ may fail to mitigate the issue of sparse gradients and local minima. The performance of \name is sensitive to changes in the variance $\sigma$. The variance $\sigma$ in \name balances the tradeoff between the convergence and the effectiveness of the optimization. Therefore, we investigate the impact of the $\sigma$ on \name's localization performance in the following experiments. As shown in Fig.~\ref{subfig:var-free}, the localization errors in device-free settings decrease with an increased variance $\sigma$. The reason is that as the excessive variance is selected, the divergence among the sampled positions degrades the performance of the gradient descent approach. 
Furthermore, in Fig.~\ref{subfig:var-based}, it can be seen that the localization errors in device-based settings first follow a similar trend as the device-free settings but reach the peak when the $\sigma$ is~0.4. After that, the localization errors start to rise at the value of 0.5. The underlying reason for the phenomenon is that device-based localization exhibits more local optimality compared to device-free configurations. Therefore, setting an appropriate $\sigma$ is more critical in device-based setups.

\subsubsection{Impact of Bias $\alpha$}\label{sec:bias}
\begin{figure}[h]
    \vspace{-2ex}
	\captionsetup[subfigure]{justification=centering}
		\centering
		\subfloat[Device-free settings.]{
        \hspace{-3ex}
		  \begin{minipage}[b]{0.49\linewidth}
		        \centering
			    \includegraphics[width = \textwidth]{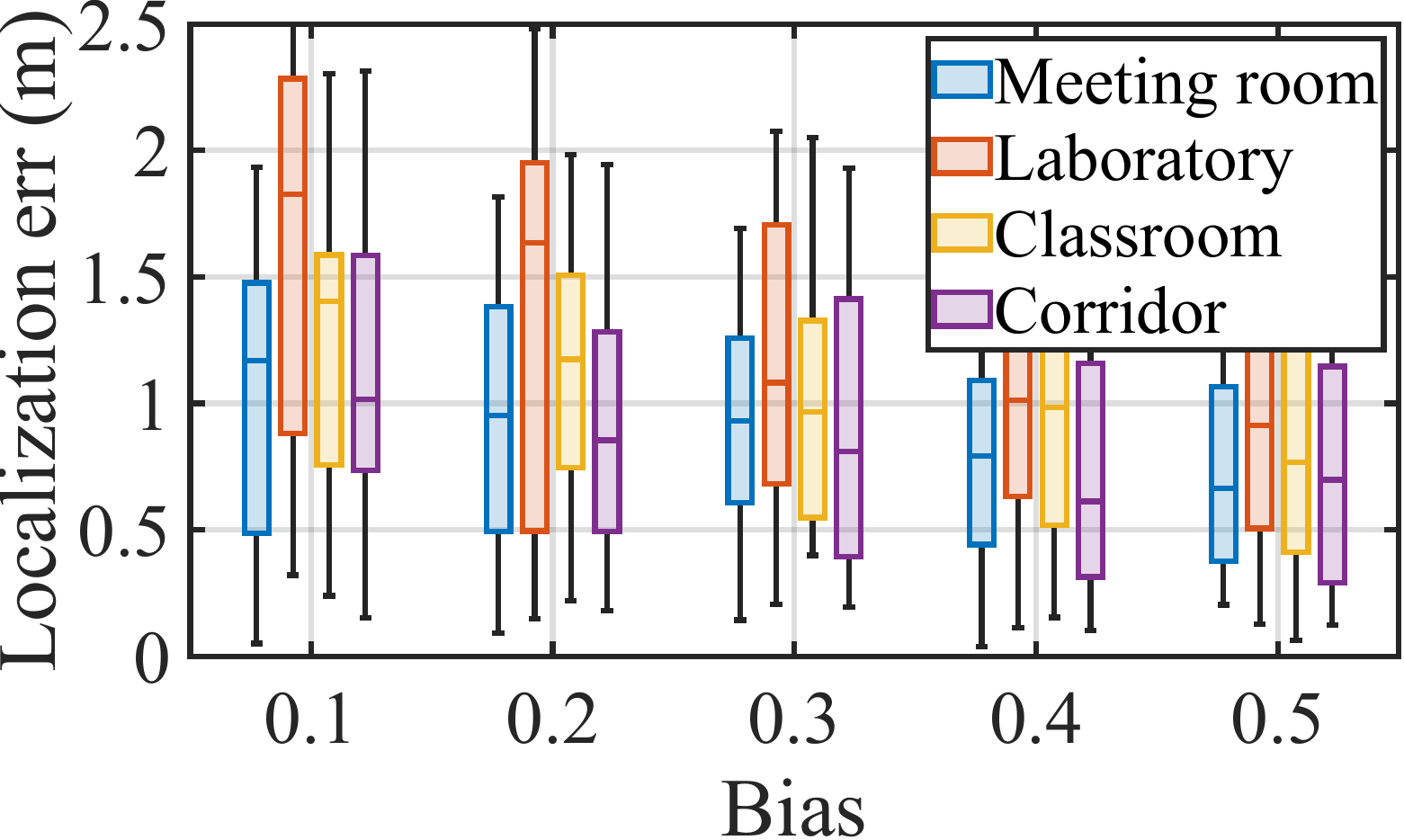}                   
			\end{minipage}
			\label{subfig:bias-free}
		}
		\subfloat[Device-based settings.]{
		    \begin{minipage}[b]{0.49\linewidth}
		        \centering
			    \includegraphics[width = \textwidth]{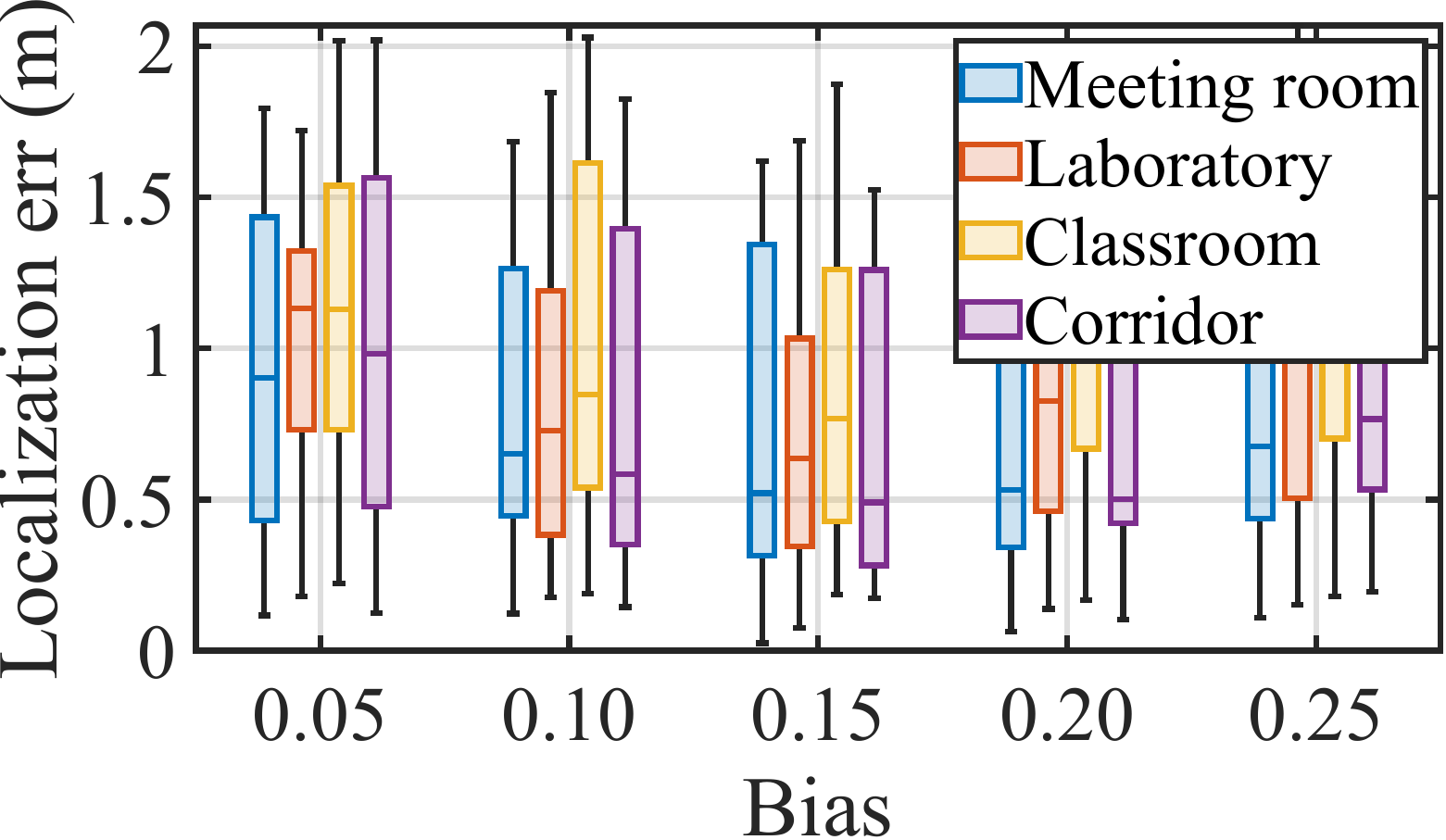}
                    
			\end{minipage}
			\label{subfig:bias-based}
		}
		\caption{Impact of bias $\alpha$.}
        \vspace{-3ex}
		\label{fig:bias}	
\end{figure}

As mentioned in \S~\ref{sec:grad-enh}, the bias parameter $\alpha$ determines the preference of \name for samples with lower loss. The $\alpha$ is a key parameter for the localization performance of \name.
The quality of the weighted gradient is highly impacted by the variations of the loss function around the current position. 
The variation is governed not only by the variance $\sigma$ but also by the bias $\alpha$. A high value of $\alpha$ makes the weighted gradient concentrated in a few sample positions leading to lower loss. This results in a degraded quality of the gradient. However, a lower value of $\alpha$ means that the gradient cannot provide sufficient focus to the position with lower loss, thereby overlooking more valuable descent directions. Therefore, an appropriate $\alpha$ value is important. We evaluate the impact of $\alpha$ on the \name performance in the section. As Fig.~\ref{subfig:bias-free} shows, the localization errors in device-free settings decrease as $\alpha$ increases to~0.5. This is because the loss is relatively small in the device-free settings. Therefore, a larger bias is preferred to better distinguish important sampling points. However, Fig.~\ref{subfig:bias-based} shows the localization errors in device-based settings first decrease as $\alpha$ increases to~0.15, and then rise. This is because, in a device-based configuration, the loss is relatively large. Therefore, a smaller bias can focus on more samples, but it should not be too small.

\subsection{Ablation Study}
\begin{figure}[h]
    \vspace{-2ex}
	\captionsetup[subfigure]{justification=centering}
		\centering
		\subfloat[Device-free settings.]{
        \hspace{-3ex}
		  \begin{minipage}[b]{0.49\linewidth}
		        \centering
			    \includegraphics[width = \textwidth]{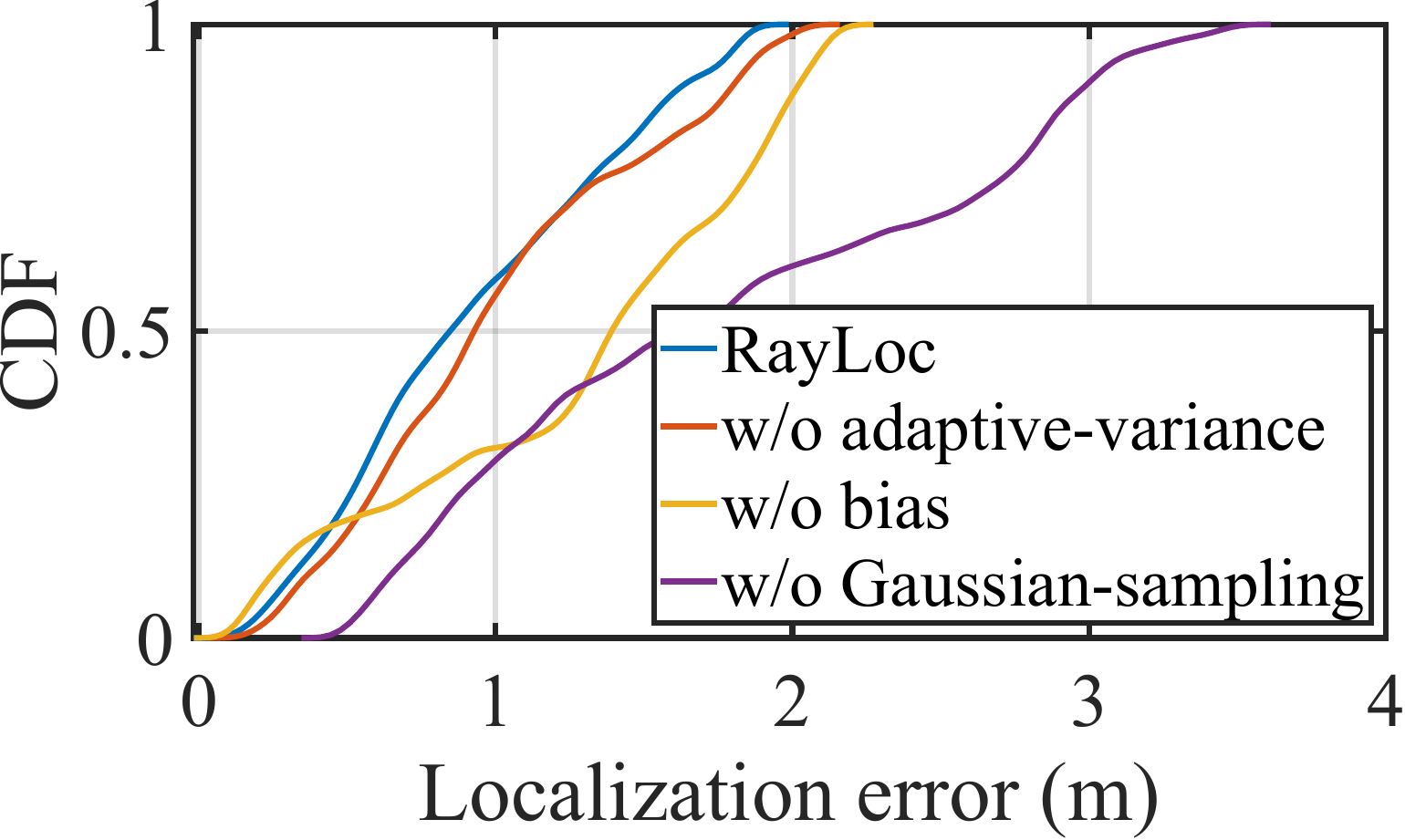}                   
			\end{minipage}
			\label{subfig:abl-free}
		}
		\subfloat[Device-based settings.]{
		    \begin{minipage}[b]{0.49\linewidth}
		        \centering
			    \includegraphics[width = \textwidth]{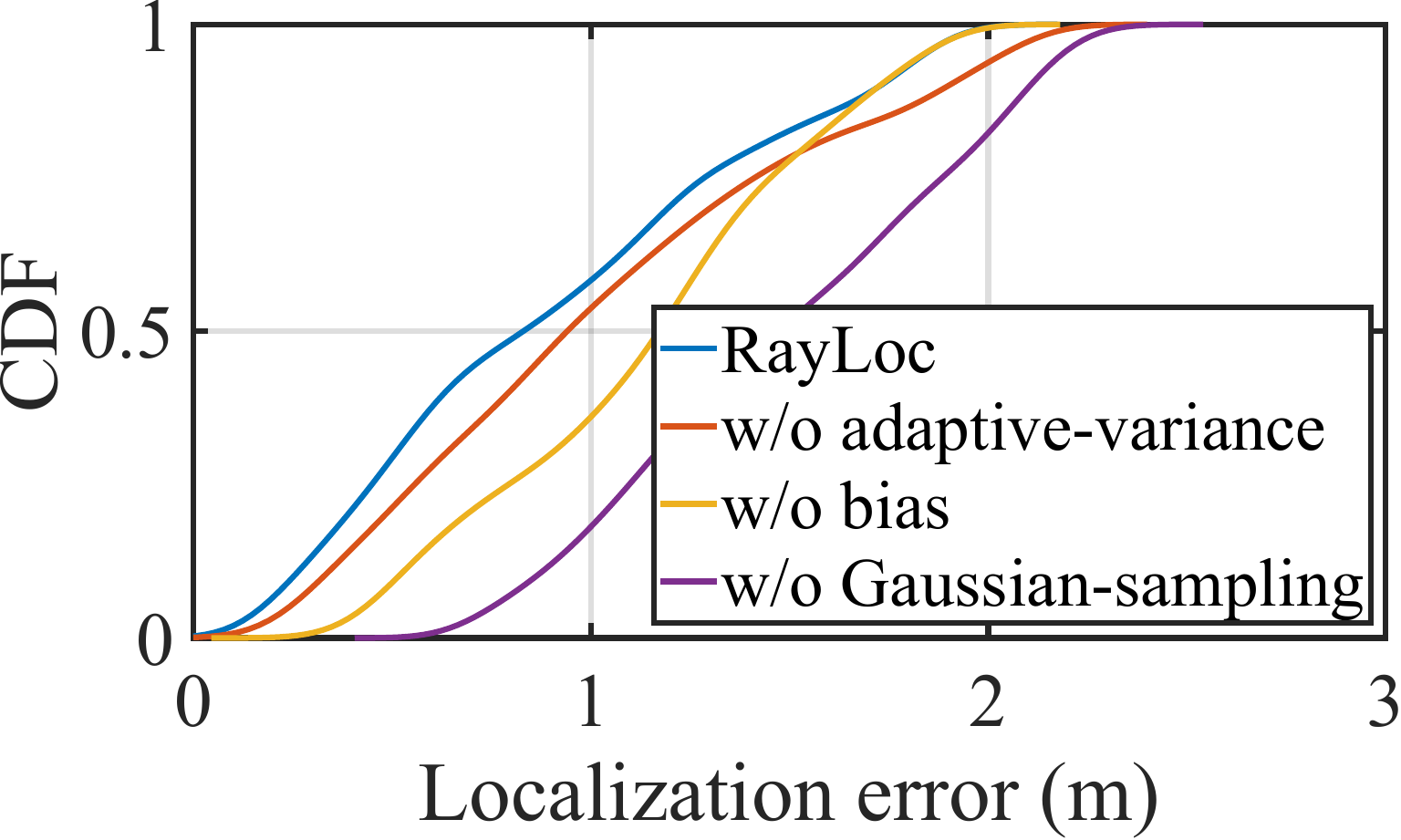}
                    
			\end{minipage}
			\label{subfig:abl-based}
		}
		\caption{Ablation study.}
        \vspace{-3ex}
		\label{fig:abl}	
\end{figure}

In this section, we conduct the ablation study to evaluate the impact of each module of \name on the localization performance. We compare the original \name effectiveness with the following cases: removing the adaptive variance, removing the bias, and removing the Gaussian smoothing. We hold all other parameters (the bias $\alpha$, the initial variance $\sigma$, and the number of sampled positions) fixed and run the same number of optimization iterations. 
The results for device-free and device-based settings are shown in Fig.~\ref{subfig:abl-free} and Fig.~\ref{subfig:abl-based}. For \name, removing the Gaussian smoothing has the most significant impact on the effectiveness of the localization. From Fig.~\ref{fig:abl}, we can clearly observe that the median localization error exceeds 1.75~\!m without the Gaussian smoothing in device-free settings while the \name maintains a media localization error of 0.67~\!m. Under the device-based setting, similar issues also arise.
This is because the presence of plateaus and local minima renders gradient-based localization schemes incapable of obtaining accurate gradient navigation, thereby preventing them from identifying the correct ground-truth location. 

\section{Discussion of Future Works}\label{sec:discussion}

\subsection{Reduction of Computational Overhead}
In this paper, we utilize physical-based RT to discover the presence of plateaus and local minima in the loss landscape and investigate the feasibility of using gradient-based methods for localization. Despite \name demonstrating effectiveness and robustness in localization, the physics-based RT can incur substantial computational overhead. To reduce computational costs, we can use a learning-based approach to control the number of emitted rays, achieving a balance between localization accuracy and computational complexity.
Additionally, we can reduce the complexity of the scene in Blender by lowering the number of triangular primitives. By doing so, we can effectively decrease the computational overhead without affecting the object's shape. In addition, we will explore neural-based RT simulation~\cite{winert} by employing a surrogate neural network to replace the physically-based RT, making \name more practical.

\subsection{Alternative Scene Model Construction Methods}
In this paper, we propose an innovative data-driven approach to constructing high-fidelity scene models by refining coarse-grained models derived from floor plans. While our method achieves impressive results, it currently requires extensive CSI measurements, which poses significant implementation challenges in real-world scenarios. To address this limitation, we envision replacing the labor-intensive model construction process with a streamlined one-shot approach, substantially reducing both measurement and computational overhead. A promising direction lies in leveraging  LiDAR and depth sensors to generate detailed 3-D models in a single pass. Looking ahead, our research will explore the enhanced capabilities of these sensors to further improve model fidelity while making scene construction more efficient and scalable across diverse applications.

\section{Conclusion}\label{sec:conclusion}
Taking an important step towards spatial intelligence, we have proposed \name for high-performance wireless indoor localization. Leveraging the inverse process of fully differentiable RT, \name is able to estimates the locations of both objects and transceivers, thereby unifying both device-free and device-based localization, previously deemed as two distinct research problems, into a single framework. Moreover, \name also enhances localization accuracy by refining scene parameters and improving optimization tractability through convolution with a Gaussian kernel. With extensive real-world experiments and baseline comparisons, we have demonstrated the promising performance of \name in wireless indoor localization.

\bibliographystyle{IEEEtran}
\bibliography{main}
\balance
\vfill

\end{document}